\documentclass[aps,twocolumn]{revtex4}

\input epsf
\usepackage{amsmath}
\usepackage{amsfonts}
\usepackage{amssymb}%
\usepackage{graphicx}

\newcommand\makeAfig[3]{\begin{figure}[htbp]  \begin{center}  \leavevmode
\vskip 0.3in
\includegraphics[width=#1 \textwidth]{png.files/#2.png} 
\end{center}
\vskip -0.00in  \caption{#3}  \label{fig:#2}
\end{figure}}

\newcommand\makeFfig[3]{\begin{figure}[htbp]  \begin{center}  \leavevmode
\vskip 0.3in
\includegraphics[width=#1 \textwidth]{png.files/#21.png} 
\includegraphics[width=#1 \textwidth]{png.files/#22.png} 
\includegraphics[width=#1 \textwidth]{png.files/#23.png} 
\includegraphics[width=#1 \textwidth]{png.files/#24.png} 
\includegraphics[width=#1 \textwidth]{png.files/#25.png} 
\includegraphics[width=#1 \textwidth]{png.files/#26.png} 
\end{center}
\vskip -0.00in  \caption{#3}  \label{fig:#2}
\end{figure}}

\newcommand{\bel}[1]{\begin{equation}\label{#1}}
\newcommand{\ee}{\end{equation}}

\newcommand{\reffig}[1]{Fig.~\ref{fig:#1}}

\newcommand{\beq}{\begin{eqnarray}}
\newcommand{\eeq}{\end{eqnarray}}

\def\be{\begin{equation}}
\def\ee{\end{equation}}

\newcommand{\rem}[1]{}

\def\pip{\pi^\wedge_v}
\def\pipm{\pi^{\wedge\vee}_v}
\def\pim{\pi^\vee_v}
\def\MET{$\hskip0.01 in{\slash}\hskip -0.09 in{E_T}$}

\begin{document}

\title{On the Phenomenology of Hidden Valleys with Heavy Flavor}
  
\author{Matthew J. Strassler}
\affiliation{Department of Physics and Astronomy,
Rutgers University, Piscataway, NJ 08854}

\begin{abstract}
A preliminary investigation of a large class of Hidden Valley models
is presented.  These models are more challenging than those
considered in arXiv:0712.2041 [hep-ph]; although
they produce a new light resonance which decays
to heavy standard model fermions, they exhibit no light dilepton
resonance.  A heavy
$Z'$ decaying to v-hadrons, which in turn decay mainly to bottom
quarks and tau leptons, is considered; six case studies are
investigated, using a new Monte Carlo simulation package.  It is found
that the one-to-one correspondence of jets and partons is badly
broken, and the high-multiplicity heavy-flavor signal probably cannot
be isolated by counting jets, with or without heavy-flavor tags.
Instead, other measures, such as counting and correlating vertices or
displaced tracks, and possibly counting of (non-isolated) muons and
use of event-shape variables, should be combined with scalar
transverse energy and/or missing transverse energy to reduce
backgrounds.  Within the resulting sample, searches for the v-pion
mass resonance in both di-jet and single-jet invariant mass can help
confirm a signal.  The best observable in a perfect calorimeter seems
to be single-jet invariant mass for jets of larger radius ($R$=0.7),
although this needs further study in a realistic setting.  A more
detailed signal-to-background study is needed as a next step, but will
face the difficulty of estimating the various high-multiplicity
backgrounds.

\end{abstract}

\maketitle

\section{Introduction}
 
The ``Hidden Valley'' scenario \cite{HV1}, if realized in nature, may
result in unusual and little-studied phenomena at the LHC.  In
this scenario, the standard model is accompanied by a nearly hidden
sector containing light particles (a ``hidden valley''). These
particles cannot currently be abundantly produced, due typically to an
energetic barrier or a weak coupling.  The increased energy of the LHC
may greatly enhance their production.  The same barrier can be
traversed in the opposite direction to allow some of these particles
to decay visibly to standard model particles.  A schematic
illustration of such models is shown in \reffig{HVdiagram}.  Examples
of hidden valley models include the original illustrative classes of
models given in \cite{HV1}, along with quirk and squirk models
\cite{TwinHiggs,KLN, HV1, FoldedSUSY} and a wide class of
``unparticle'' models \cite{Un1} with an added mass gap
(e.g. \cite{unFox, unQuiros}), whose signals were discussed in detail
in \cite{HVun}.  Another related class of examples was studied in
\cite{SchabWells,BowenWells, GW}.  Motivation for such sectors is
provided by, among other possibilities, supersymmetry-breaking models,
which often introduce one or more hidden gauge groups.  While these
gauge groups are normally imagined to be unimportant at LHC energies,
this reflects a theoretical bias.  Such sectors might also be
responsible for dark matter, and may have an important role to play in
other aspects of particle physics \cite{EspQuir}, astrophysics  and
cosmology \cite{JMR}.

The main interest of these models for the LHC era is that their
signatures are often distinctive, and can differ from the many
supersymmetry, little Higgs, extra-dimensional and technicolor
signatures that have been so often discussed.  These include
high-multiplicity events (generally non-thermal and non-spherical),
possibly with large missing energy, and exhibiting large
event-to-event fluctuations.  New decay modes for Higgs bosons
\cite{HV1, HV2,HVun}, supersymmetric particles \cite{HV3,HVun}, and
top quarks \cite{HVun} often arise.  Light neutral resonances are
common, lighter perhaps than 100 GeV or even 10 GeV.

A common hidden valley signature, but one which I will {\it
not} address in this paper, is displaced vertices.  In certain regions
of parameter space, some of the new light particles have long lifetimes,
decaying at macroscopic distances.  There is no standard model background
to compute or estimate.
Key issues associated with such signals involve experimental
challenges: triggering, detector noise, beam halo, pion collisions
with detector material, vertex reconstruction, {\it etc.}  All of these are
detector-specific, and any study of this signature requires a full detector
simulation.

However, it may happen that all the new particles decay promptly, or
that some decay promptly and all others are stable and invisible.  In
this case, identifying the resulting high-multiplicity and often
low-rate signal, over a large standard model background,
becomes a challenge that can be addressed in part through theory 
and simulation.  Below I will consider models of this type.

In some classes of hidden valley models \cite{HV1,HVun}, there is a
new and frequently-produced particle that often decays to electron and
muon pairs.  It is then relatively easy to discover the signal, as
emphasized in \cite{HVWis}.  Simple and rather crude event-shape cuts
that remove the largest backgrounds may enhance signal-to-background
to the point that, using the excellent low-$p_T$ dilepton mass
resolution of the LHC experiments, a resonant peak could be detected.

In this paper, I will consider a much more difficult situation.  I
will examine a class of hidden valley models {\it with prompt decays
and heavy-flavor final states, and with no dilepton resonances}.  Some
of these models also have large missing energy.  All have large
event-to-event fluctuations in the multiplicity of standard model
partons in the final state.  The background to such signals is
difficult to estimate, because it consists of a cocktail of many
different processes, none of which can be calculated beyond leading
order in $\alpha_s$, and few of which can be identified and measured
using the data itself.  Because of this, it will be a considerable
challenge to carry out a concrete signal-to-background study.  The
goal of this article is to lay the groundwork for such a study, and
suggest features of the signal which could be used in any search
for a hidden valley of this type.

First I will outline the specific model that is chosen as an exemplar
from this class, and will describe the Monte Carlo event generation
package used to study it.  Then I will describe the case studies,
examining the basic phenomenological features of the signal, in
Sec.~\ref{sec:casestudies}.  After exploring the basic underlying
phenomena, I will consider how jets are constructed in the signal.
Finding that jets are not fully sufficient for interpreting or
isolating the signal, I will consider other non-standard methods for
reducing background.  Finally, in Sec.~\ref{sec:resonance}, I will
examine the question of how to identify the resonance whose
observation would confirm the signal, considering both dijet and
single jet invariant mass.  A summary of results and some additional
comments are given in the conclusion; two appendices fill in some details
on secondary muons and on jet algorithms.
\makeAfig{0.32}{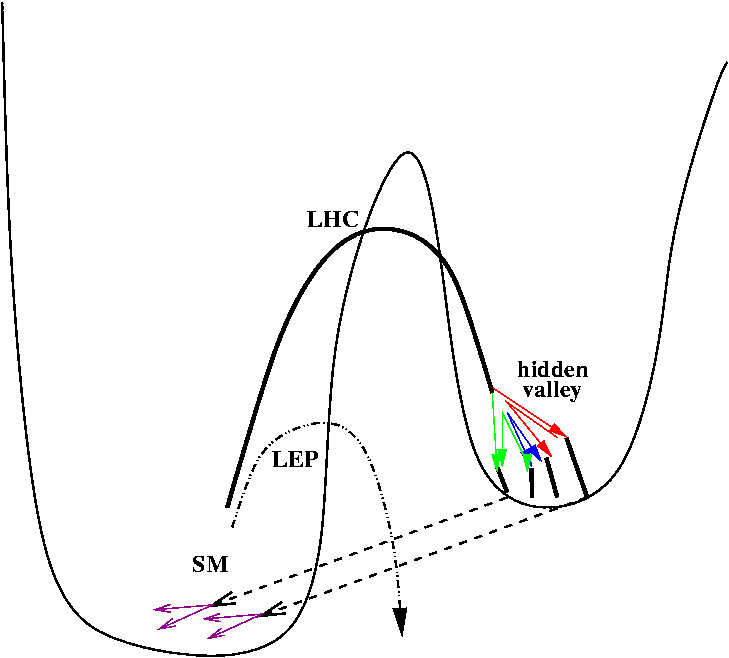}{A schematic illustration of models in the
hidden-valley scenario.  With sufficient energy, available at the LHC
but not at LEP, a barrier may be traversed that allows production of
new light states in a hidden sector.  Dynamics in the hidden sector
may produce large numbers of particles.  Some of these new particles
may decay back to the standard model, often with long lifetimes.}

\subsection{The Models}
\label{subsec:models}

Several large classes of hidden valley models share the phenomenology
of high-multiplicity final states, rich in heavy flavor and possibly
missing energy.  These include a wide variety of confining hidden
sectors whose light stable hadrons are all pseudoscalar and/or scalar
mesons with comparable masses; an example was given in \cite{HV1}.
Another class involves weakly-coupled models with multiple electroweak
doublet and singlet Higgs bosons which mix together.  These models
have been discussed widely (see \cite{unQuiros} and references
therein) but their potential for high-multiplicity heavy-flavor final
states was only recently recognized \cite{CFW,HV2,HVun}.  A third
class can include strongly-interacting hidden valleys which couple to
the standard model mainly through the Higgs boson; these have not yet
been explored fully.

In this paper I will consider the theory in \cite{HV1}, as a very
simple example from the first category.  This is a hidden valley which
closely resembles QCD.  To make this study especially straightforward,
I have chosen a hidden valley sector (``v-sector'') that, like QCD, has an $SU(3)$
gauge group and two light ``v-quarks'' $U$ and $D$, with masses adjusted
so that the light ``v-hadron'' mass ratios are those of QCD.  It is important
to emphasize that
this model is a stand-in for a much larger class of models.  Indeed
there is no reason for the physics of a hidden valley to closely
resemble QCD, any more than technicolor models should closely resemble
QCD.  However, for initial studies of v-sector phenomenology, the case
of a QCD-like v-sector is simplest to investigate first.  This is
because the physics is easy to understand, and a Monte Carlo event
simulator is easily constructed.

As in \cite{HV1}, where more details are given, I will consider such a
v-sector coupled to the standard model through a broken $U(1)$ gauge
symmetry, under which both standard model particles and the v-quarks
carry a charge.  The $Z'$ gauge boson of the $U(1)$ will serve to
mediate both production and decay of particles in the v-sector.
\makeAfig{0.22}{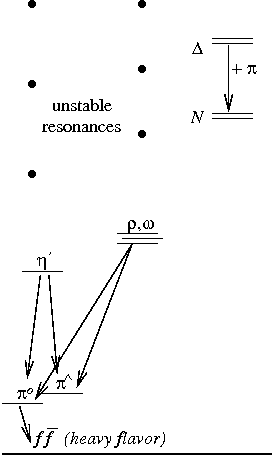}{The spectrum of the v-sector considered in
this paper.  In analogy to QCD, all v-hadrons rapidly decay down to v-pions
and v-nucleons; then the $\pi_v^0$ (and in the B cases, also the
$\pip$ and $\pim$) decay more slowly to standard model fermion pairs,
preferentially to heavy flavor.}

The long-lived v-hadrons of a QCD-like v-sector with two light
v-quarks $U$ and $D$ are three light v-pions and a heavier v-nucleon
doublet, as shown in \reffig{2LFspect}.  All other v-hadrons (such as
the v-rho and v-Delta) decay immediately to v-pions and v-nucleons.
For simplicity, it is assumed that v-baryon number is conserved, so
the v-nucleons are stable and invisible.  The three v-pions $\pi_v$, a
triplet under v-isospin, consist of a v-flavor off-diagonal pair with
quantum numbers of $U\bar D$ and $D\bar U$, analogous to the
$\pi^{\pm}$ of QCD, and a third, the v-flavor diagonal v-pion with
quantum numbers of $U\bar U-D\bar D$, analogous to the $\pi^0$.

A point of notation: it is natural to name the v-pions as
($\pi_v^\pm$, $\pi_v^0$), in analogy to QCD's pions ($\pi^\pm,
\pi^0$).  This notation was used in \cite{HV1}.  However, the use of
the $\pm$ superscript proves confusing, because {\it all the v-pions
are electrically neutral} --- after all, they are part of a hidden
sector.  To avoid any confusion of this type (and at the expense of
introducing another), I will call the $U\bar U-D\bar D$ state
$\pi^0_v$, but call the $U\bar D$ state $\pip$, and the conjugate
$D\bar U$ state $\pim$.  \makeAfig{0.281}{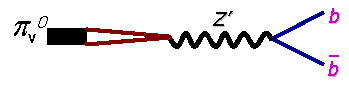} {The $\pi_v^0$
decays via a $Z'$ to heavy flavor.  The $\pipm$, if unstable, decays
through a v-flavor-changing interaction to the same final state.}

If the third component of v-isospin $I_v^3$ is conserved, then the
$\pip$ is stable and invisible, but the breaking of total v-isospin
allows the $\pi_v^0$ to decay via a $Z'$ back to standard model
particles, as shown in Fig.~\ref{fig:vpidecay_new}.  Helicity suppression
assures the spin-zero $\pi_v^0$ decays mainly to heavy fermions (for
the same reason that $\pi^+\to\mu^+\nu$ decays dominate over $\pi^+\to
e^+\nu$ in QCD); branching fractions are roughly proportional to
squares of fermion masses.  In the particular model of \cite{HV1}, and
for light v-pion masses, the width of the $\pi_v^0$ is
\begin{equation}\label{Gpitotlo}
\Gamma_{\pi_v^0} \sim
6\times 10^{9} {\rm \ sec}^{-1}{f_{\pi_v}^2 m_{\pi_v}^5\over
(20\  {\rm GeV})^7}
\left({10\ {\rm TeV}\over  m_{Z'}/g'}\right)^4  \ .
\end{equation}
which has a very strong dependence on model parameters; here $f_{\pi_v}$
is the v-pion decay constant, while $m_{Z'}$ and $g'$ are the $Z'$
mass and coupling.    
\makeAfig{0.366}{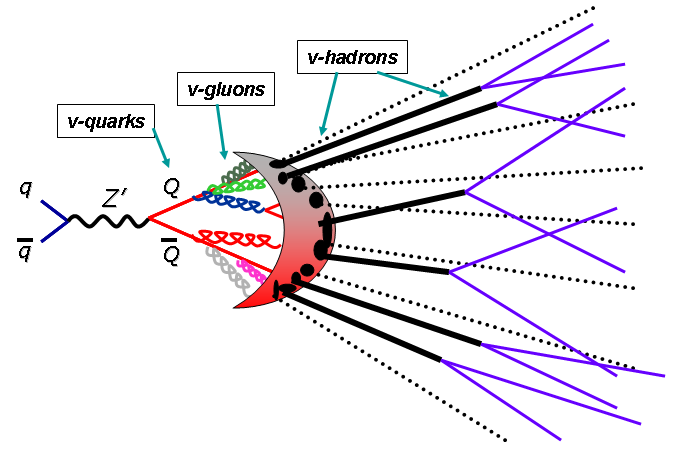}{A $Z'$ decays to two v-quarks, which emit
v-gluons in a v-parton shower.  These then are confined into v-pions
and v-nucleons.  Some of the v-hadrons 
(shown dotted) are stable and invisible, but
others are metastable and decay, mainly to $b\bar b$.}

It is also possible that the third component of v-isospin $I_v^3$ 
is violated.
In this case even the $\pip$ and $\pim$ can decay, with widths that
are smaller than that of the $\pi_v^0$ by a factor which is a
dimensionless measure of $I_v^3$ breaking.  In this article, I
simply assume that {\it either} (A) $I_v^3$ is conserved (so that
the $\pip$ is stable and invisible) {\it or} (B) $I_v^3$ is badly
violated (so that the $\pip$ decays promptly.)   The case studies
will be divided into ``A cases'' and ``B cases'' according to this
distinction.

The basic production process for these particles is shown in
\reffig{Zp2XXYY_new}.  It involves $q\bar q\to Z' \to Q\bar Q$, where
$Q$ is a v-quark.  The v-quarks undergo a parton shower through
v-gluon emission, following which they are confined by strong
v-interactions into v-hadrons.  These v-hadrons decay down to
v-nucleons and v-pions, and some of the v-pions may then decay visibly
to standard model particles.
\makeAfig{0.336}{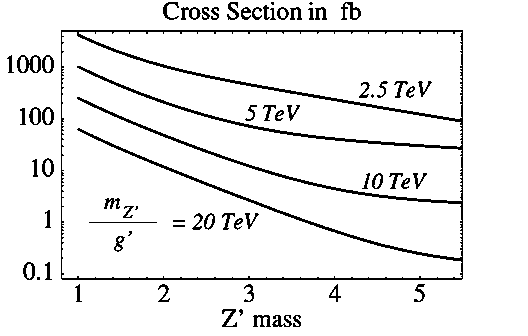}{Cross-section for v-particle production via a
$Z'$ in the QCD-like model of \cite{HV1}.  The cross-section in other
models
may easily differ by
an order of magnitude; see text.}

In the study below, I will consider a $Z'$ with mass 3.2 TeV.  The
v-pion masses will range between 50 and 200 GeV.  In this case the
v-pions decay promptly, and the production cross-section is expected
to be of order 10--100 fb in the model of \cite{HV1}, see
\reffig{cross}.  In other models the cross-section could be different
by a factor of 10 or so, larger or smaller, due for example to
different $Z'$ charge assignments, or to a different number of colors in
the v-sector.

In summary, the model considered below has gauge group $[SU(3)\times
SU(2)\times U(1)]_{SM} \times U(1)'\times SU(3)_v$, with the $U(1)'$
broken at the few TeV scale, the $SU(3)_v$ group confining at the few
hundred GeV scale, and two light v-flavors of v-quarks $U$ and $D$.
Standard model fermions and v-quarks all carry some charge under the
$U(1)'$, allowing the $Z'$ to serve as a communicator between the two
sectors.


\rm

\subsection{The Hidden Valley Monte Carlo 0.5}
\label{subsec:HVMC}

The event simulator HVMC 0.5, upon which all studies in this paper are
based, is described in this section.  (A more general Monte Carlo
simulator has been developed with S. Mrenna and P. Skands
\cite{HVMC1}, and studies based upon it will be presented elsewhere.)
HVMC 0.5 \cite{HVMC0} is built on existing tools, which are rather
easy to modify for current purposes.  In particular, elements of
PYTHIA \cite{Pythia} are strung together to simulate a v-sector which
is isomorphic to three-color two-flavor QCD, with all masses and other
dimensional quantities scaled up, relative to QCD, by a constant
factor $R$.  \makeAfig{0.366}{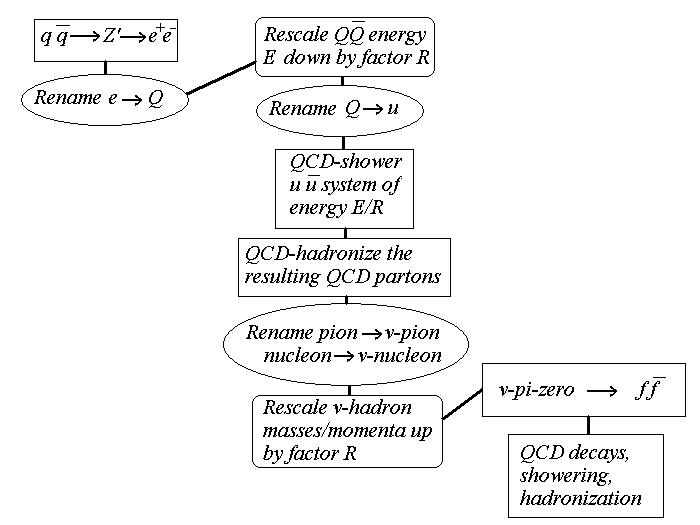}{The algorithm used in HVMC 0.5.}

A v-sector with three colors, two flavors and confinement scale
$\Lambda_v$ has v-pions with mass $m_{\pi_v}=m_\pi R$, where $R\equiv
\Lambda_v/\Lambda_{QCD}$, and $\Lambda_{QCD}$ is the QCD confinement scale.
It also has nucleons with similarly scaled-up masses.  The $\eta'_v$ (the
iso-singlet pseudoscalar of 2-flavor QCD) has its mass set to $m_\eta
R$.  Then, given a mass $M$ for the $Z'$, the simulation of events
proceeds as in \reffig{HVMC}.
\begin{itemize}
\item The process $q\bar q\to Z' \to Q\bar Q$ is simulated, where $q$
is an ordinary quark and $Q$ is a v-quark, using the PYTHIA routine for
$q\bar q\to Z'\to f\bar f$.
\item The v-parton showering and v-hadronization of the $Q\bar Q$ system
is simulated.  This is done by
\begin{itemize} 
\item scaling down the energy of the $Q\bar Q$ system from its original
energy $E_0$ to the energy $E=E_0/R$;
\item simulating QCD parton showering
and hadronization (with the number of light flavors set to 2)
of an ordinary quark-antiquark system with center-of-mass energy equal
to $E$;
\item scaling up the masses and momenta of the 
final-state 
QCD hadrons (with $\pi^0$'s undecayed) by the factor $R$, and renaming
them as v-hadrons.
\end{itemize}
\item The decay of the v-hadrons to standard model partons is simulated
using PYTHIA decay routines.
\item The decays, showering and hadronization of the standard model
partons and the simulation of the underlying event proceed using the
usual PYTHIA routines.
\end{itemize}
The resulting final states consist of standard model hadrons, photons and
leptons, along with stable neutral v-hadrons that escape undetected.
{\it All results in this paper are
based on analysis of the final state hadrons, photons and leptons without
accounting for detector effects, other than geometric acceptance,
except where otherwise noted.}

For the sake of clarity (though the actual effect on the studies below
is small) it should be noted that two-flavor QCD as simulated in this
way is not quite a consistent model.  The tuning of PYTHIA to match
existing data on hadronization, branching fractions, {\it etc.}, is
not correct for a two-flavor model.  The iso-singlet would-be
Nambu-Goldstone boson $\eta_v$ is now affected by the anomaly and
takes the place of the $\eta'$.  There are small effects on the
nucleon mass from the slight differences in the running coupling that
are similarly ignored.  But these issues are of minor impact on the
phenomenology and of minor concern for the current studies.  As there
is no reason to expect the hidden sector in nature to be of exactly
the form considered here, the aim of this paper is not precision but
rather phenomenological and experimental guidance, in search of robust
analysis strategies.
\makeAfig{0.35}{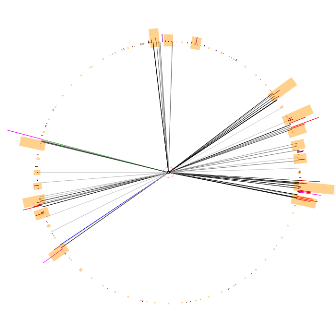}{A schematic view of a typical event from
case A1.  The view is along the beampipe. Charged tracks are shown in
the ``tracker'', the central disk in the figure.  Since there is no
magnetic field in this event display, all tracks with $p_T<3$ GeV have
been removed; grey-level corresponds to $p_T$, with hardest tracks
shown in black.  Neutral hadrons and photons are indicated as
outward-pointing lines starting at the outer edge of the tracker, and
calorimeter energy in azimuthal angular bins of width $2 \pi/60$ are
shown as bars at the outer edge of the tracker.}

\section{The Case Studies}
\label{sec:casestudies}

\subsection{Preliminaries}
\label{subsec:prelim}

The studies below will all involve decays of a $Z'$ of mass 3.2 TeV to
a hidden valley sector.  In many models the $Z'$ will already been
discovered in its decays to dilepton final states.  However, knowledge
of its presence and of its mass do not significantly aid in uncovering
the hidden valley signal, because of the latter's complexity.  In other
models, the branching fraction of the $Z'$ to dileptons will be too small,
and the $Z'$ will not yet have been identified when the hidden valley signal
is sought.

As noted in Sec.~\ref{subsec:models}, signal cross-sections in the 10--100 fb
range are consistent with LEPI and LEPII constraints.  The
cross-section is easily changed by an additional factor of 10, without altering
the observable phenomenology in any other way, by adjusting the
$U(1)'$ coupling constant $g'$ (see \reffig{cross}).  For simplicity,
I will study samples of 1000 events, such as might be obtained in a
real LHC analysis.  {\it All plots shown below, except where noted,
  show results for 1000 simulated signal events.}  \
\begin{table}[b]
\begin{tabular}[c]{||c||c|c|c|c|c|c|c||}\hline
\ Case & 
$\pip$ & $m_{\pi_v}$ & $R$ &
\# $\pi_v$
& $\hat H_T$ & $M_4$ & $\slash \hskip -0.1 in E_T$ 
\\
\ & stable?      & (GeV)       &   & decays
& (GeV) & (GeV)& (GeV) \\
 \hline 
 A1 & Yes & $50$ &  $368$ & 4.0 & 667 & 590 & 318 \\
 A2 & Yes & $120$ & $883$ & 2.4 & 765 & 667 & 400 \\
 A3 & Yes & $200$ &  $1470$ & 1.5 & 886 & 770 & 459 \\
 B1 & No  & $50$ &  $368$  & 10.3 & 1650 & 1427 & 214 \\
 B2 & No  & $120$ & $883$  & 6.1 & 1835 & 1562 & 182 \\
 B3 & No  & $200 $ & $1470$ & 3.9 & 2248 & 1810 & 145 \\
\hline 
\end{tabular}
\caption{The case studies, showing the stability of the $\pip$, the
mass of the v-pion, the ratio $R=\Lambda_v/\Lambda_{QCD}$, the
average number of visible v-pion deays, and the average $\hat H_T$,
$M_4$, and \MET.}
\label{table:cases}
\end{table}

The case studies are distinguished by the masses of the v-pions and by
whether the $\pip$ are stable or decay promptly.  In Table
\ref{table:cases} the cases are listed.  In addition to the masses and
decay settings, the table shows the average multiplicity of
visibly-decaying v-pions and some kinematic information: the
transverse calorimeter energy $\hat H_T$, the
invariant mass $M_4$ of the four highest-$p_T$ jets, 
and the average missing transverse
momentum (MET, or \MET).
The quantities $\hat H_T$ and \MET\ are computed here using scalar and
vectorial sums of the $p_T$ of all calorimeter towers,
\begin{equation}\label{HTdef}
\hat H_T \equiv \sum_{towers} \left|\vec p_T\right| 
\Theta\left(|\vec p_T|-5\ {\rm GeV}\right)\ 
\Theta\left(|\eta|-3\right)\ ;
\end{equation}
\begin{equation}\label{METdef}
\slash \hskip -0.1 in E_T \equiv \left |\sum_{towers} \vec p_T\right| \ .
\end{equation}
The calorimeter towers combine the 3-momenta of the various
hadrons, electrons, photons {\it and muons} in $0.1\times 0.1$ bins in
pseudorapidity $\eta$ and azimuthal angle $\phi$.  For $\hat H_T$ I
include only towers with $p_T>$ 5 GeV and $|\eta|<3$, as indicated by
the $\Theta$ functions, to reduce significantly the impact of the
underlying event.  (However, for a fully realistic study, an $H_T$
variable built from the reconstructed jets might be much more robust;
in these models, a variable $H_T^{(jet)}$, defined as the scalar
summed $p_T$ of all jets with $p_T>25$ GeV and $|\eta|<3$, takes
values about 10 percent larger than $\hat H_T$ defined above.)
Meanwhile $M_4$, the invariant mass of the four highest-$p_T$ central
jets, is built from jets defined using the midpoint cone algorithm of
cone radius 0.4; see Sec.~\ref{subsec:jets} and Appendix
\ref{app:jets} below.  
\makeAfig{0.38}{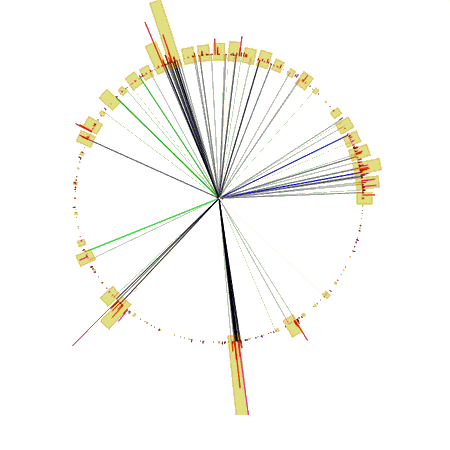}{As in
\reffig{R050_30_evt}, a schematic view of a typical event from case B1.}

Note the obvious progressions in the table.  Comparing A1, A2 and A3,
one sees the decrease in the multiplicity of v-pions; the same trend
appears in B1, B2 and B3.  Meanwhile the B cases, with a decaying
$\pip$, have roughly triple the number of visibly decaying pions, much
higher visible energy, and much less \MET, compared to the A cases for
the same v-pion mass.  For illustration, event displays of one event
each from cases A1, B1 and A3 are given in
Figs.~\ref{fig:R050_30_evt}--\ref{fig:R200_32_a}.  However the reader
should bear in mind that event-to-event fluctuations in appearance are
much larger here than in most standard model backgrounds or
traditional new signals such as gluino production.
\makeAfig{0.35}{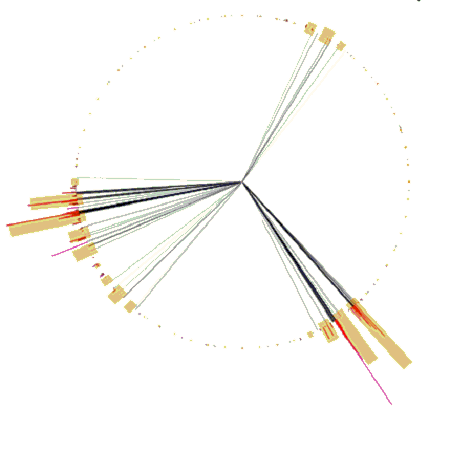}{As in \reffig{R050_30_evt}, a schematic view
of a typical event from case A3.}


\makeAfig{.366}{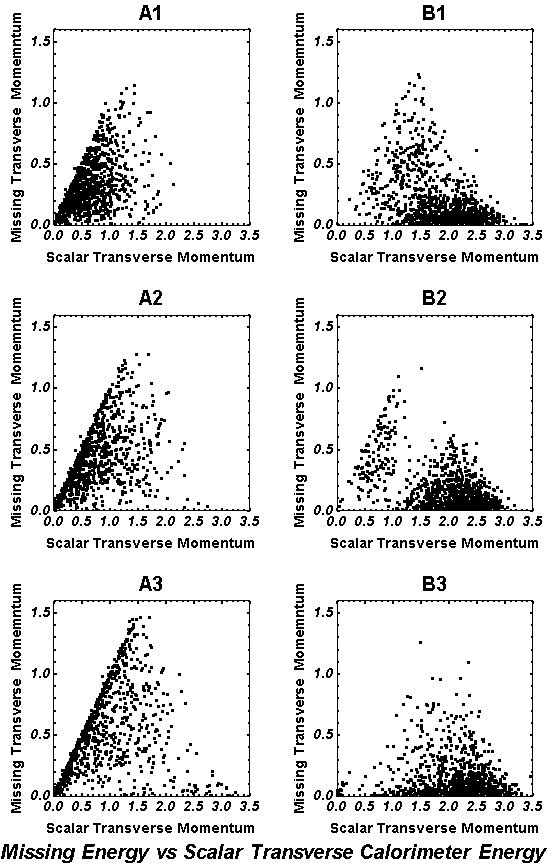}{The distribution of missing transverse
momentum (MET) versus $\hat H_T$ in TeV.  These quantities
are defined in Eqs.~(\ref{HTdef})-(\ref{METdef}).}  Figure \ref{fig:xHTMET} shows $\hat H_T$
versus \MET\ for the various cases.  The difference between the A
cases, where roughly 2/3 of the v-pions are stable and escape undetected, and
the B cases, where the v-pions all decay promptly and most of the
debris from the $Z'$ is observed, is obvious.  Occasional v-baryons
(stable and invisible in all present case studies)  can
provide some \MET\ even in the B cases.  There is additional and
sometimes substantial \MET\  from secondary
neutrinos produced in semileptonic decays of 
$b$ and $c$ quarks and especially in $\tau$ decays.

Already from these plots, one sees clearly that {\it the A cases will
have much larger standard model backgrounds than the B cases.}  The B
cases are more similar to those studied in \cite{HVWis}, though with a
higher invariant mass, lower rate, and no dilepton resonance.  I will
show in Sec.~\ref{subsec:jets} that they have many reconstructed hard
jets.  Backgrounds are high-multiplicity QCD events with many $b$
quarks, including $t\bar t b\bar b$, $b\bar b b \bar b jj$, $t\bar t
t\bar t$, $t\bar t Z$, {\it etc.}  These signals tend to be in the few pb
range or less, and will be greatly reduced by an $\hat H_T$ cut at,
say, 800 GeV.  The A cases, by contrast, despite their large \MET,
are often in the same overall kinematic regime
as relatively low-energy standard model processes with a few jets and
\MET.  These much larger backgrounds include $W$ or $Z$ plus jets
(especially heavy flavor), $t\bar t$ plus jets, $t\bar t W$ or $t \bar
t Z$, $t\bar t b\bar b$, {\it etc.}  Cuts on $\hat H_T$ and \MET\ that have
high efficiency for the signal will still leave considerable amounts
of background behind.


In contrast to standard model backgrounds, triggering should not be a
problem for either the A or B cases.  Most of the events in this
signal will pass the various jet(s) or jet(s)-plus-\MET\ triggers,
with the latter being most efficient for the A cases.  Even in the A
cases, due to secondary muons, many events will also pass dimuon and
muon-plus-\MET\ triggers; see Appendix \ref{app:leptons}.  The trigger
will mainly remove unspectacular, low-visible-energy events, which are
rare in the B cases and consist of a large minority in the A cases.
But the events that fail these triggers are precisely those which
would be especially difficult to distinguish from standard model
background off-line.  Conversely, the events which are most
distinctive --- with multiple acoplanar high-$p_T$ jets and possibly
large \MET\ --- will be among those which will pass the trigger.  {\it
For this reason, the effect of the trigger is likely to be relatively minor,
compared to the other issues addressed below.}
\makeAfig{.336}{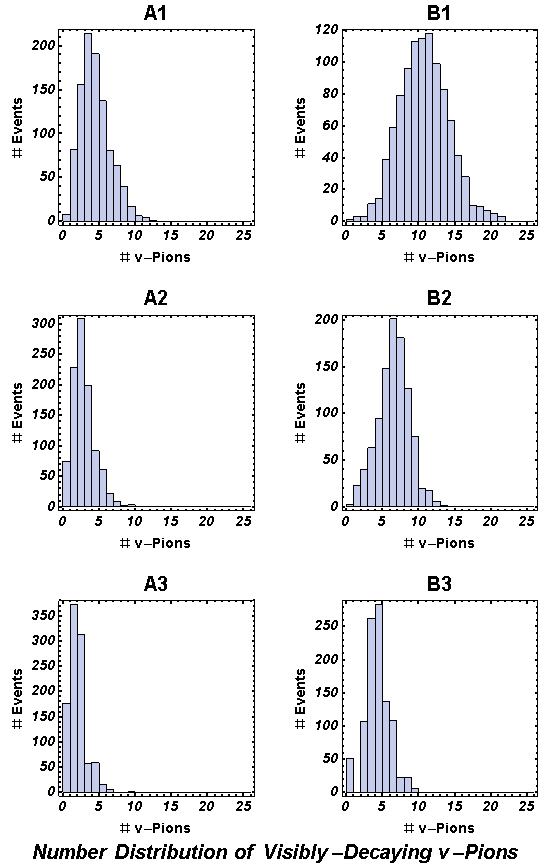}{For the case studies, with 1000 events,
the distribution of the number of visibly-decaying v-pions, each decaying to two standard model particles.}
\makeAfig{.366}{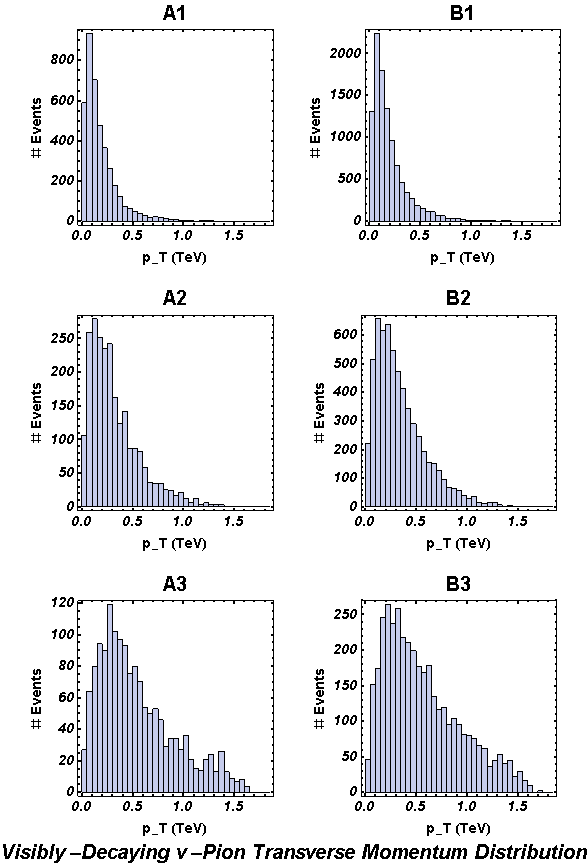}{As in the previous figure, the $p_T$ distribution
of the visibly decaying v-pions, in TeV.}
\makeAfig{.336}{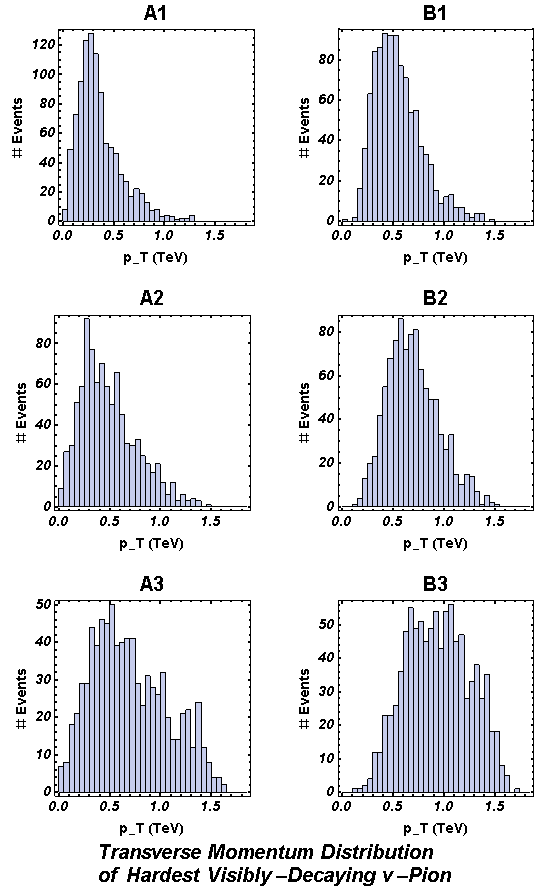}{As in the previous figure, the $p_T$ distribution 
of the {\it hardest} ({\it i.e.}, highest-$p_T$) visibly decaying v-pion.}

\subsection{The v-pions}
\label{subsec:vpions}

In these case studies, the $Z'$ decay produces a substantial number of
v-pions, organized into two rather fat v-jets.  All of these v-pions
decay visibly in the B cases, while about a third are visible in the A
cases.  (Actually the fraction is a bit larger in these QCD-like
models, due to v-isospin violation that, as in QCD, biases the decay
of the $\eta_v$ toward $\pi_v^0$s.)  The number of visible v-pions is
shown in \reffig{xNvpivis}.  Since the only difference between case B1
and case A1 is the stability of the $\pipm$, the distribution of
visible v-pions in case B1 equals the distribution of all v-pions,
visible and invisible, in case A1.  The same applies for B2 and A2,
and for B3 and A3.

 In \reffig{xvisvpipT} are shown
the $p_T$ distributions of the v-pions; notice few have $p_T<50$ GeV,
and most are relativistic.  \reffig{xhivispT} shows the $p_T$
distribution of the highest-$p_T$ visibly decaying v-pion, which is
almost always relativistic, often with a boost factor above 3.
This is relevant because {\it the majority of events have a
highly-boosted v-pion whose decay products are separated in $\eta$ and
$\phi$ by $\Delta R < 0.5$.}; here $\Delta R=\sqrt{(\Delta
\eta)^2+(\Delta\phi)^2}$ as usual.  We will see in
Sec.~\ref{subsec:jets} and Sec.~\ref{sec:resonance} that the decay
products of high-$p_T$ v-pions often are merged into single jets.

As an aside, let us note that the multiplicity distribution and $p_T$
distribution of the v-pions, and the distribution of final-state quark
and lepton flavors, is quite different from a thermal distribution, as
one would (at least naively) expect in a black hole decay
\cite{GidThom,Lands}, another potential source of high-multiplicity
events.  Also the events are not in general spherical (see
Figs.~\ref{fig:R050_30_evt}--\ref{fig:R200_32_a}) in constrast to expectations
for black holes.  (See however Sec.~IV D of \cite{HVun}.)

\subsection{v-Pion Decay Products}
\label{subsec:daughters}

As noted in Sec.~\ref{subsec:models}, the decay rates of the $\pi_v^0$
(and the $\pipm$ if unstable) to quarks and leptons are roughly
proportional to the square of the final-state fermion masses.  The
fermion mass used should be evaluated at the v-pion mass scale.  For
simplicity the relative branching ratios to quarks and leptons are
taken to be those of a Higgs boson of the same mass as the $\pi_v$.
Since the $\pi_v$ does not decay (at tree level) to $WW$ and $ZZ$,
these decay channels are first removed before the branching fractions
are computed.  In the mass ranges considered here, the unstable
v-pions decay mainly to $b\bar b$ pairs with a large branching
fraction ($\sim 90\%$ for the lighter v-pions), with the remainder
going mainly to $\tau^+\tau^-$, $c\bar c$ and gluon pairs.  The number
of final-state (short-distance) standard model particles is simply
double the number of visibly-decaying v-pions.  Note from Table
\ref{table:cases} that the average multiplicity of final-state partons
ranges from 3 in case A3 to 20 in case B1.  Also, note in
\reffig{xNvpivis} the wide fluctuations and the long tail, which
reaches 42 in case B1 and even in case A3 extends to 12.  (However, a
fraction of these partons have low transverse momentum, as will be
seen below.)  The event-to-event fluctuations in the multiplicity of
final-state partons are exceptionally high compared to most
new-physics signals.  This is part of what makes this signal
challenging.  \makeAfig{.336}{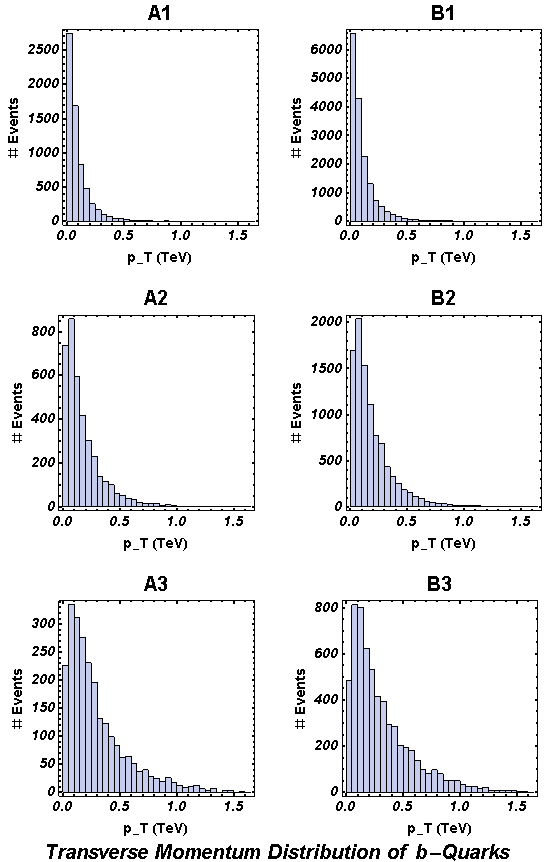}{For the case studies, over 1000
events, the $p_T$ distribution of the daughters (mainly $b$ and $\bar
b$ quarks) of the v-pions.}

\subsubsection{Bottom Quarks and Taus}
\label{subsubsec:bstaus}

Most of the final state partons, especially for lighter v-pions where
decays to gluons are suppressed, are $b$ quarks and antiquarks.
The $p_T$ spectrum of the central ($|\eta|<2$) $b$ quarks is
shown in \reffig{xbpT}.  For lighter v-pions, the higher multiplicity of
$b$ quarks is somewhat compensated by their lower $p_T$, which makes
them less likely to produce jets above kinematic cuts and to decay
with detectable vertices.  The fraction of bottom quarks that are
central ($|\eta| < 2$) and hard ($p_T>$ 50 GeV) varies from about
55\% for cases A1 and B1 to nearly 85\% for cases A3 and B3.

Taus are produced in roughly ten percent of the v-pion decays, and are
common in these events.  One could imagine that central hadronic taus
could be useful in identifying this signal.  However, in the present
studies, few events have more than two tau leptons, which is not
enough to be unusual (given $t\bar t$ rates).  In addition, one or
both tau's will sometimes fail isolation requirements, either because they
are too close together (as we will see below) or because of the high
multiplicity environment in which they are produced.  For this reason,
the number of taus identified is likely to be too small
for it to play a role in extracting
the signal.  But it should be noted that in other
hidden valley models, where the $\tau$--to--$b$ ratio might be
enhanced, the role of taus in signal identification might be more
important.

\begin{table}[b]
\begin{tabular}[c]{||c||c|c|c|c|c|c||}\hline
\  & A1 & A2 & A3 & B1 & B2 & B3 \\
 \hline 
Fraction of events   &&&&&&\\
with 3 or more muons & 0.09 & 0.04 & 0.02& 0.46 & 0.28 & 0.11 \\
($p_T>3$ GeV, $|\eta|<2.5$)&&&&&&\\
\hline
\end{tabular}
\caption{Fraction of events with multiple muons.
All cases studies have
1000 events.}
\label{table:features}
\end{table}

\subsubsection{Secondary Muons and Electrons}
\label{subsubsec:mue}

In these models, unlike those of \cite{HVWis}, electrons and muons do
not provide a direct handle for discovering the v-hadrons directly.  The
branching fraction of the v-pion to muons is tiny (unless the v-pion
is lighter than $2 m_b$, in which case it will be very long lived).
However, because these events have high multiplicity, and because the
v-pions decay mainly to $b$, $c$ and $\tau$, which in turn can produce
light leptons (generally non-isolated), it is very common for one or
more electron or muon to be produced as a secondary.  The presence of
these light leptons could assist with reducing backgrounds.

Since the leptons in question are not typically isolated, I focus on 
muons, which are easier to identify.  The numbers of events
with three or more muons that have $p_T>3$ GeV and $|\eta|<2.5$, with
no isolation requirement, are shown in Table \ref{table:features}.
Note the large number in the B cases.  This implies that requiring multiple muons
may serve as one of several useful criteria for selecting events for
analysis.  Unfortunately the number of muons is too small
to be of much use in the A cases.

Additional plots related to lepton distributions appear in Appendix \ref{app:leptons}.
\makeAfig{.336}{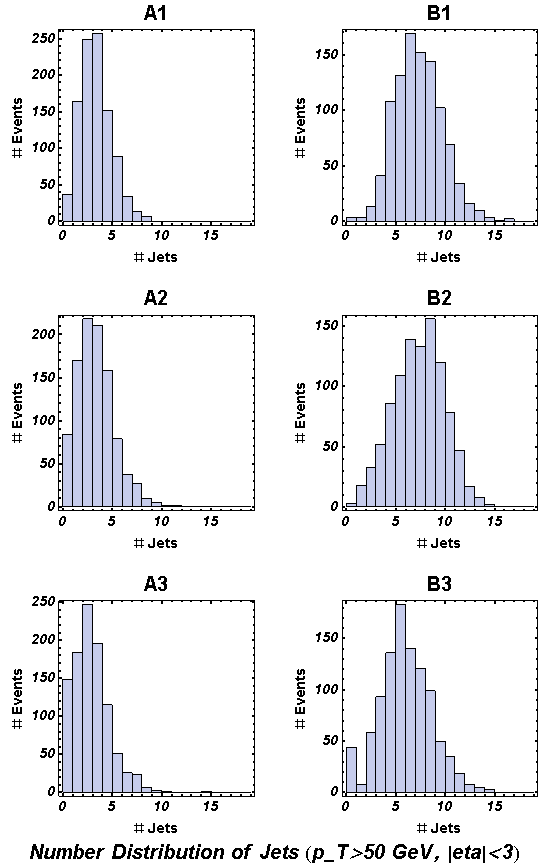}{For the case studies, with 1000 events, the
number of jets (formed using 
the midpoint cone algorithm with
cone-size $\Delta R=0.4$; see text for more details) with $p_T> 50$
GeV and $|\eta|<3$.  The number of jets is considerably smaller than
twice the number of visibly-decaying v-pions, \reffig{xNvpivis}.}

\makeAfig{.366}{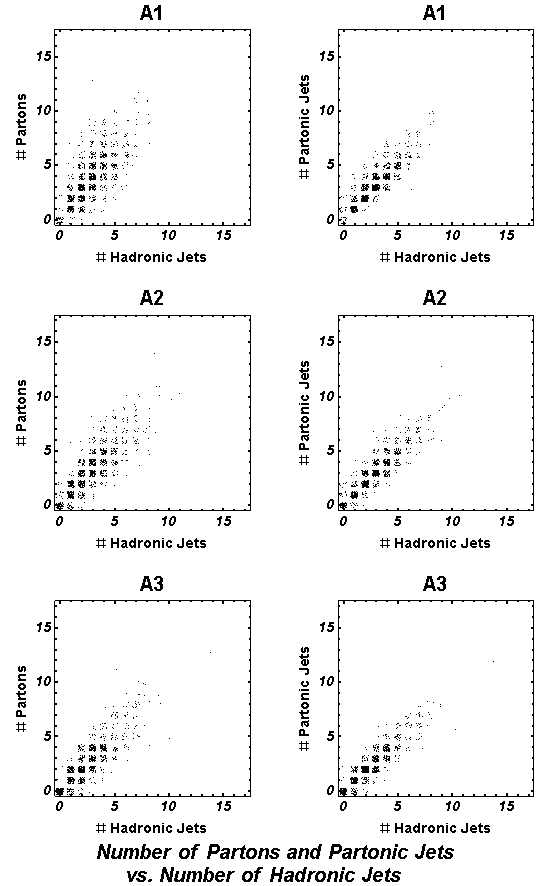}{For the A cases, with 1000 events, the number
of partons versus the number of hadronic jets (left plot) and the
number of partonic jets versus the number of hadronic jets (right
plot).  Jets are formed using the midpoint cone algorithm with
cone-size $\Delta R=0.4$; see text for more details.  Cuts of $p_T>
50$ GeV and $|\eta|<3$ are imposed on both partons and jets.  One sees
that that partonic and hadronic jets are in correspondence, but partons do
not correspond as well to jets.}

\makeAfig{.366}{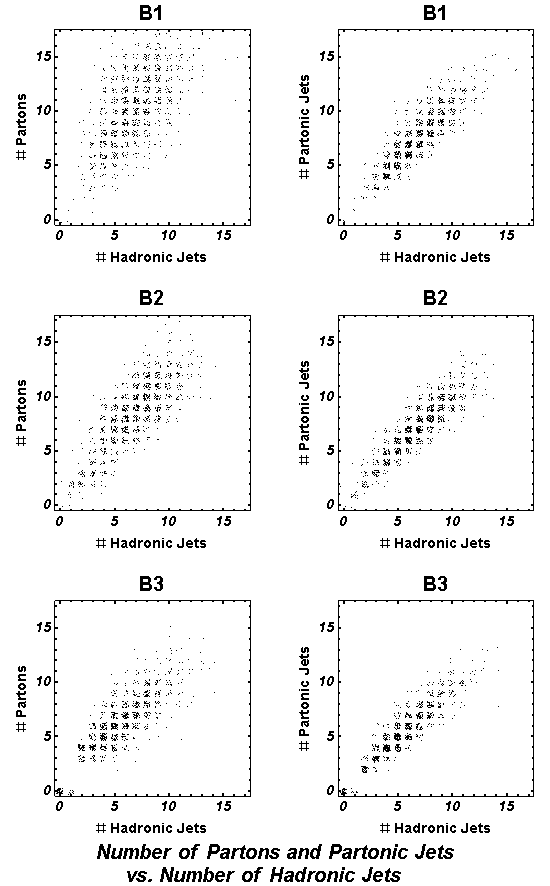}{As in the previous plot, for the B cases.}

\subsection{Jets}
\label{subsec:jets}

Typically one characterizes events on the basis of
``objects'', where the objects include electrons, muons, photons,
hadronic taus, and jets, which may be tagged or untagged.  Since the
majority of the many jets in the signal are from $b$'s, one might expect
about half of them on average to be tagged, with a few events
containing an exceptional number of tags.  One might expect these
events to be the ones that stand out above standard model background.

This expectation is not entirely wrong, but it is also too naive.  In
particular, {\it the standard jet-parton correspondence does not work
in this signal.}  As we will see, the number of well-reconstructed and
taggable jets is typically considerably smaller than the number of $b$
quarks.  Many jets contain two or more $b$ quarks.  While the tagging
efficiency may be somewhat higher in such jets, that the number of tagged jets 
obviously cannot be exceptional if the number of
jets itself is not exceptional.  

Instead, here and in Sec.~\ref{subsec:beyondjets}, I will argue that
{\it treating jets as objects, characterized as either ``tagged'' or
``untagged'', would throw away crucial information needed to separate
this signal from background.}  A substantial fraction of the jets in
this signal are not standard jets, and it appears that this fact may
be critical in suppressing backgrounds.

In the plots shown in the main part of this article, the midpoint-cone jet
algorithm is used, with cone size $0.4$; more details on the
parameters chosen are given in the appendix.  Changing parameters, or
choosing other algorithms, will change the details, but as argued in
Appendix \ref{app:jets}, will not change the main conclusions of this
section.  The same may not be said for finding the v-pion resonance,
however; see Sec.~\ref{sec:resonance}.

In \reffig{xNjets} is shown the number distribution of jets per event
with $|\eta|<3$ and $p_T>50$ GeV.  The average number of jets is
rather large but not spectacular in the B cases, and not very large in
the A cases, which do not even have a substantial tail on the high
side.  This means one cannot find this signal by simply demanding
large numbers of hard jets; in the A cases, a requirement of more than
six jets removes most of the signal, and preserves only half of the
signal in the B cases.  Recall that these signals are in the 10--100
fb range, whereas multi-$b$ backgrounds with 8 or more jets, from
$t\bar tb\bar b$, $t\bar tZ$, {\it etc.}, are in the 1--10 pb range.  A hard
cut on the number of jets cannot be afforded.

Note also that in all cases the average number of jets is
significantly lower than twice the average number of visibly decaying
v-pions, shown in \reffig{xNvpivis}.  The typical
v-pion is not producing two jets.  It is important to identify the
reason for this.
\makeAfig{.47}{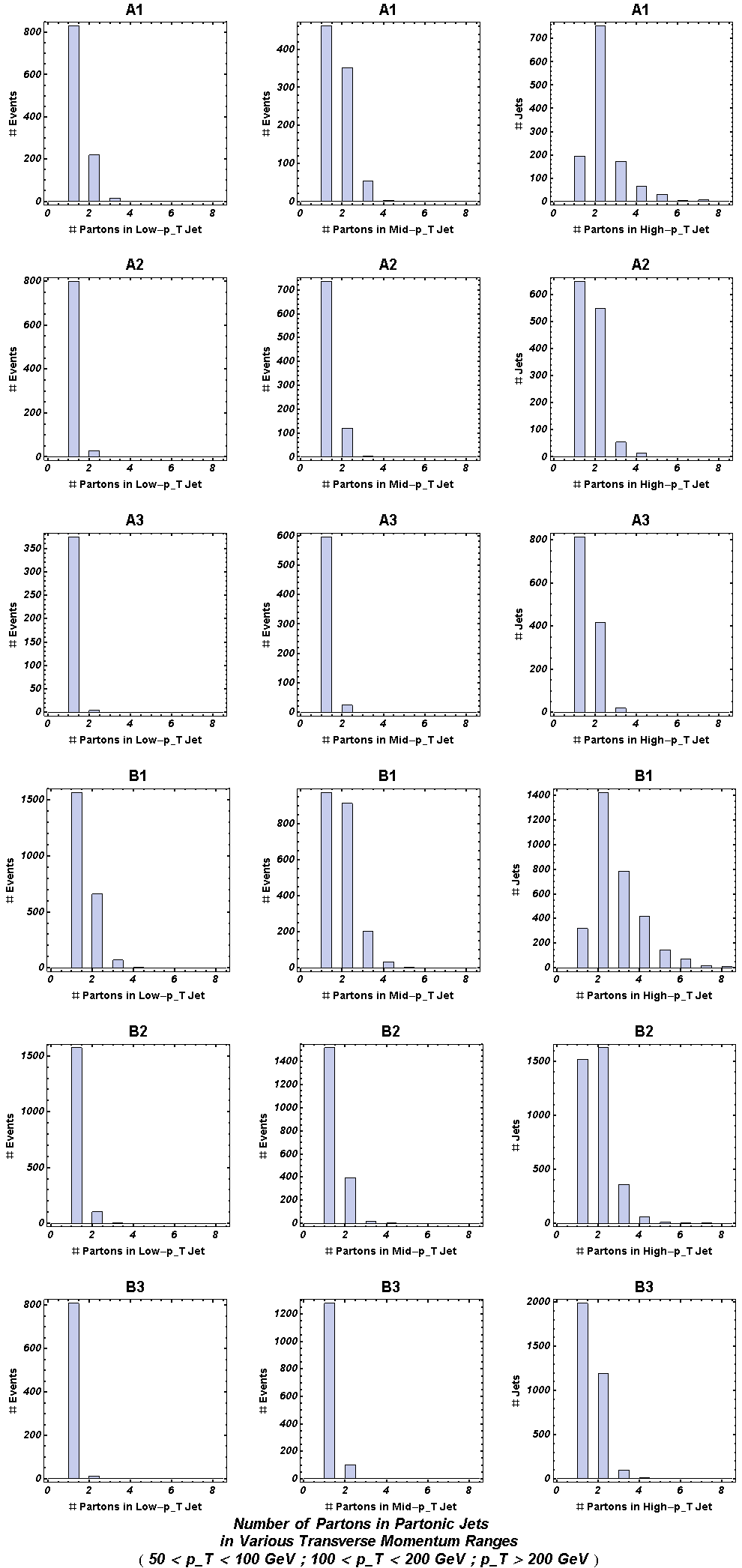}{For the case studies, with 1000 events, the
number of partons inside of partonic jets (which correspond well to
hadronic jets) for jets with $|\eta|<3$ and $p_T$ in three ranges: $50
< p_T < 100$ GeV (left plots), $100 < p_T < 200$ GeV (middle plots) and
$p_T > 200$ GeV (right plots).}

Let us first quantify the degree to which the jets do not correspond
well to the partons in the event.  (In this section, ``partons''
refers to v-pion daughters, which appear at short distance; it does
not refer to partons emerging through subsequent showering.)  That
there is a mismatch is hardly surprising, given the cluttered nature
of these high-multiplicity events.  In the left-hand plots
of Figs.~\ref{fig:xNpNpjA} and \ref{fig:xNpNpjB} 
are shown the
number of partons versus the number of jets; here $p_T>50$ GeV and
$|\eta|<3$ for both partons and jets.  Notice these often differ by as
much as a factor of 1.5 to 2.  This breakdown of the jet-parton
correspondence is natural and indeed has been seen before; it will
certainly can occur in $t\bar t t\bar t$ events, and in events with
highly boosted massive particles, such as $W$'s, $Z$'s, $h$'s and
$t$'s.  In this signal, however, it can become extreme.

Fortunately, there remains a close connection between the clustering
of hadrons and the clustering of partons.  This is partly due to the
fact that the final-state quarks are all produced in the decays of the
color-singlet v-pions, which limits the radiation of gluons at large
angles. In the right-hand plots of 
Figs.~\ref{fig:xNpNpjA} and \ref{fig:xNpNpjB} is shown
the relation between {\it jets of hadrons} and {\it jets of partons}.
Here, the hadrons are clustered according to an algorithm, the partons
({\it i.e.}, the short-distance v-pion daughters) are clustered
according to the {\it same} algorithm, and the results are compared.
Clearly the correspondence of hadronic jets and partonic jets is much
closer than that of hadronic jets and partons themselves.  In other
words, even in this signal, a fixed algorithm applied at the hadron
level gives jets that do correspond well to the jets obtained by
applying the algorithm at the parton level.  It is shown in Appendix B
that this result is robust for both cone and $k_T$ algorithms; compare
the figures above with Figs.~\ref{fig:kNpNpjA} and \ref{fig:kNpNpjB}.
In particular, although cone and $k_T$ algorithms will find different
jets in general, {\it both algorithms find the same jets in
short-distance partons as they do in hadrons.}

Since the partonic jets {\it are} a good surrogate for the hadronic
jets, one can take a short-cut to learn about the failure of the jet-parton
correspondence.  To determine precisely the number of $b$ quarks in each jet,
one should examine the $B$ mesons within the hadronic jets, but this
is technically tedious and has subtleties.  Instead, one can examine
the number of partons collected within partonic jets in various $p_T$
ranges, as shown in \reffig{xpINj}.  This provides sufficient information to illustrate the
key phenomenological points.

Several effects tend to cause multiple partons to be combined into a
single jet.  First, a high-$p_T$ jet is likely to be a single boosted
v-hadron, and so contains two partons.  This effect becomes
substantial for boost factors above about 3 or 4.  This is visible in
the right-hand plots of \reffig{xpINj}, where it can be seen that jets
with $p_T>200$ GeV have a substantial probability to contain two
partons, ranging from 1/3 for the higher-mass v-pions of cases A3 and
B3 to 3/4 for the lower-mass v-pions of cases A1 and B1.

Also, the fat v-hadronic jets from the $Z'$
decay tend to throw multiple v-hadrons into the same region of $\eta$
and $\phi$, so the probability that partons from different
v-pions are nearby in $\eta$ and $\phi$ is non-negligible. 
Furthermore, 
the presence of soft partons from the softer v-pions tends to increase
the probability that harder partons will be merged by a jet algorithm.
These combined effects can be seen in
\reffig{xpINj}, which shows that a
significant number of jets contain 3 or more partons, even as many as
$6$ or so, for the cases with lower v-pion mass and consequent
higher-multiplicity.  Of course the effect is more dramatic for the B
cases.

\begin{table}[b]
\begin{tabular}[c]{||c||c|c|c|c|c|c||}\hline
\  & A1 & A2 & A3 & B1 & B2 & B3 \\
 \hline 
Fraction of $b$ quarks &&&&&&\\
with $p_T<30$ GeV & 0.27 & 0.11& 0.05 &0.25 &0.10 & 0.04 \\
 \hline 
\# $b$ quarks per event &&&&&&\\
with $p_T<30$ GeV & 1.86 & 0.44& 0.12 &4.49 &1.02 & 0.28 \\
\hline
\end{tabular}
\caption{Average distributions over 1000 events of soft $b$ quarks.}\label{table:softbs}
\end{table}

Table \ref{table:softbs} provides some information about $b$ quarks
with $p_T$ less than 30 GeV, which often cannot generate a clean jet
and are rarely taggable.  Recall that the multiplicity distributions
in these signals have long tails, so the average number is less than
half the maximum.  This completely different effect also tends to
reduce the number of jets relative to the number of v-pion daughters.
Particularly in the case of lighter v-pions, these low-$p_T$ $b$
quarks contribute additional sources of confusion for jet
reconstruction, as well as adding tracks and neutrals in the few GeV
range but without providing a detectable vertex.

Altogether, this means that the number of jets is significantly less
than the number of partons.  This is a bit disappointing, as the high
multiplicity of partons is a unique and striking feature of the
signal.  {\it The mere counting of jets, even with heavy-flavor
tags, is unlikely to be enough to separate
signal from background, especially in the A cases.}

\subsection{Beyond Jets: Vertexing and Tracking}
\label{subsec:beyondjets}

To identify this signal, it seems likely that {tagging of individual
jets is not enough.}  By definition, the number of heavy-flavor-tagged
jets cannot be larger than the number of jets.  But the number of $B$
mesons can greatly exceed the number of tagged jets, as suggested in
Figs.~\ref{fig:xNpNpjA} and \ref{fig:xNpNpjB}.  In other words,
although these events do not have an exceptional number of taggable
jets, often four or less in the A cases, they do have an unusual
number of $B$ mesons.  {\it Thus to distinguish the signal from
background, it is essential to detect as many vertices from the $B$
mesons as possible.}

More precisely, {\it a remarkable feature of this signal is the number
of vertices and the distinctive correlations between vertices and
hadronic jets.}  Most of the high-$p_T$ jets contain more than one
$B$-meson vertex.  This can occur in some standard model backgrounds
(through boosted $h$ decays, boosted $Z$ decays, and most commonly
through $g\to b\bar b$ splitting) but the probability of having two or
more jets with multiple $B$ mesons is low, and the probability of
having additional $B$ mesons in the event is also low.  (Of course,
even a single $B$ may produce a second vertex when its daughter $D$
meson decays, but the kinematic correlations between the parent and
daughter vertex are distinctive and different from those of the two
$B$ mesons from a $\pi_v$ decay.)  It thus appears that moving beyond
``tagged jets'' and ``untagged jets'' as the basic objects of analysis
is important for separating signal and background.
\makeAfig{.336}{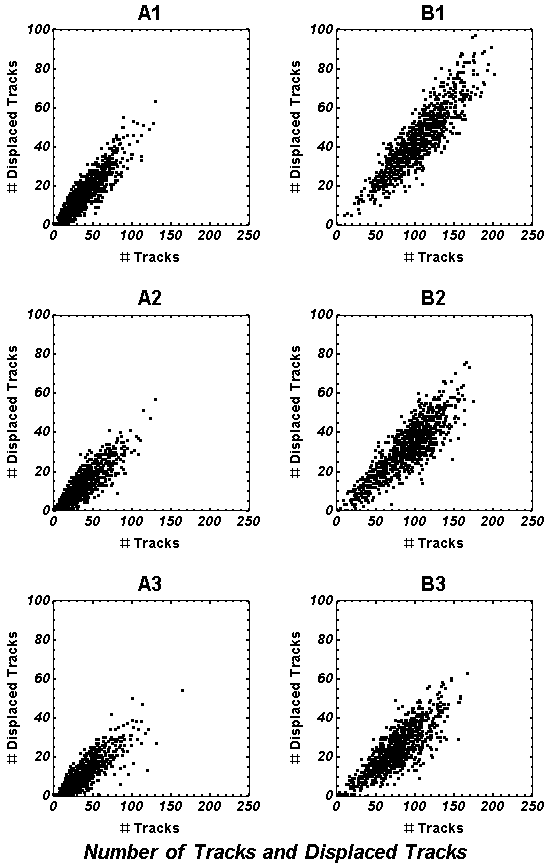}{For the case studies, over 1000 events, the
number of displaced tracks versus the total number of
tracks.  In this plot all tracks have $|\eta|<2$, $p_T>$ 2 GeV, and
displacement in three dimensions must exceed 300 microns.}
\makeAfig{.336}{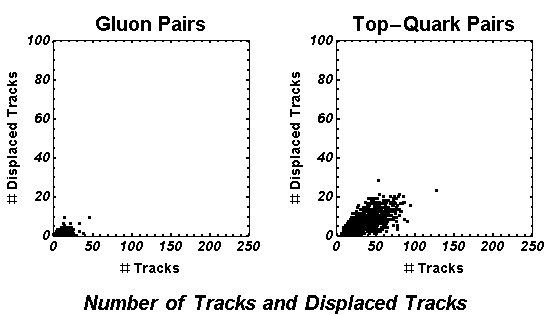}{As for the previous figure, for di-gluon
events (left-hand plot) and for $t\bar t$ (right-hand plot), generated
with PYTHIA, with radiation and the underlying event.  In both cases,
the samples have $\sqrt{\hat s} > $ 1 TeV.}

Meanwhile, vertex/jet correlations are not the only non-object-based
measure that can be useful.  Consider \reffig{xdtrks}, in which
distributions of the number of tracks with $|\eta|<2$, $p_T>$ 2 GeV,
versus the number of such tracks with three-dimensional impact
parameter $>$ 300 $\mu$m, are shown.  This is a measure of both the
number of $B$ mesons produced and the fraction of tracks that were
produced in a $B$ meson decay.  In particular, notice that the slope
(the fraction of tracks that are displaced) is somewhat larger than in
the right-hand plot of \reffig{xdtrksjt}, which shows $t\bar t$
produced at $\sqrt{\hat s}\geq$ 1 TeV.  Thus high-multiplicity
heavy-flavor events will have many tracks of which an unusually large
fraction will be displaced.  

The clustering of the displaced tracks may also be a useful variable,
which I have not yet considered.  It would be interesting to explore
a clustering observable acting upon them.

Presumably, the techniques discussed here would be
useful for many other possible new signals.  The need to move beyond
``tagged'' or ``untagged'' jets is far more general than this
particular class of models.  It should apply in any signal in which a
$b\bar b$ pair is produced by a boosted particle, such as a $Z$ or
$h$.  (For recent relevant work, see \cite{boost}.)
Obviously the number of tracks and the fraction of tracks
displaced are blunt instruments, sensitive to any process with
long-lived particles, whether $b$'s or something exotic and new.  The
clustering of tracks and vertices, however, will be variable from
signal to signal.  For instance, although in the present signal the
number of vertices is larger than the number of jets, this inequality
need not hold.  In signals with novel heavier long-lived particles,
which may decay to multiple jets at a displaced point inside the
beampipe, a number of jets may share the same vertex, and then the
number of vertices per jet may be smaller than one.  For example, were
case B3 altered so that the lifetime of the $\pipm$ were a few
picoseconds, and were the dominant decay $\pipm\to gg$, then a single
decaying $\pipm$ would make two jets, with many displaced tracks,
emerging from a single vertex.  At the other extreme, there are models
in which a large number of light long-lived states are produced, and
these can have many vertices.  Examples would include cases A1 and B1
with the v-pion mass reduced to 30 GeV and its lifetime extended to a
few picoseconds.  Then the number of vertices could be very large due
to a large v-pion multiplicity, one vertex per v-pion at the point of
its decay, and one vertex for each of the daughter $B$ mesons from the
v-pion decay.  This complex of vertexing issues deserves a thorough
exploration by the $b$-tagging community at the LHC detectors,
including LHCb.
\makeAfig{.336}{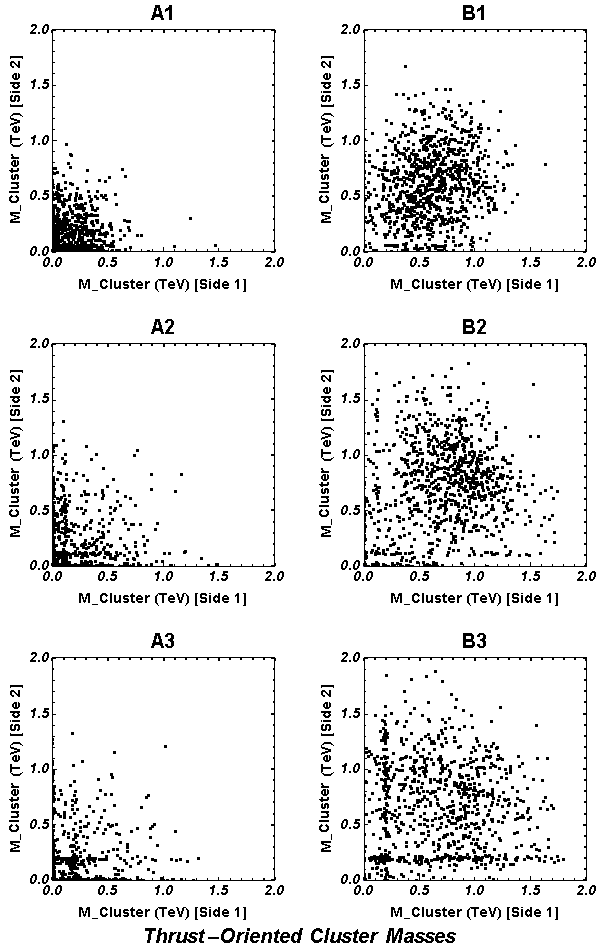}{For the case studies, over 1000 events, the
(jet-level) cluster mass in the two hemi-cylinders divided along the
transverse thrust axis.  Cuts of $p_T>25$ GeV and $|\eta|<2$ are
imposed on the jets.  See cautionary remarks in the text.}
\makeAfig{.336}{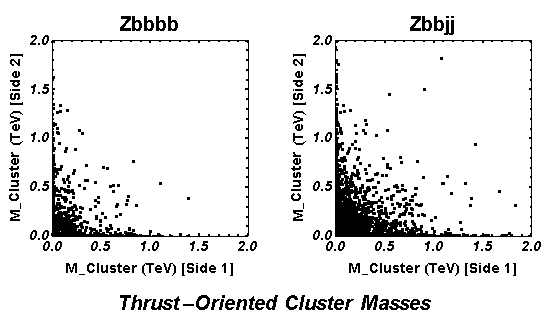}{The (parton-level) cluster mass
distribution for the $Zb\bar b b \bar b$ background ($\sigma\sim 10$
fb, 4200 events shown) and $Zb\bar bjj$ background ($\sigma\sim 2$ pb,
10000 events shown).  See text for more details and cautionary remarks.}

\subsection{A Comment on Event-Shape Variables}
\label{subsec:shapes}

Here I will briefly explore an event-shape variable for the signals
and for two backgrounds.  In \cite{HVWis} it was suggested
that transverse thrust (defined in the two dimensional plane
transverse to the beamline) is a useful variable for separating signal
and background.  Another variable used was ``cluster mass'', obtained
by dividing the cylindrical detector along the plane perpendicular to
the transverse thrust axis into two hemicylinders, and computing the
invariant mass of all activity within the hemicylinder.  To reduce the
impact of the underlying event and of initial state radiation, only
calorimeter cells satisfying certain $p_T$ and 
$\eta$ cuts were used.  

In \cite{HVWis}, the presence of a dilepton resonance makes even a low
signal-to-background ratio acceptable.  
Here, the signal is smaller, and a much better signal-to-background
ratio is needed for the signal to be confirmed (through the methods of
the Sec.~\ref{sec:resonance}).  This requires that all cuts have high
efficiency for the signal.  There are important multi-jet backgrounds
(with and without \MET\ for the A and B cases respectively) that
can be disregarded in the case study of \cite{HVWis}, but cannot be
ignored here.

A proper background study needs to account for many background
processes.  Below we consider only two, for illustration.  These are the
$\sim$10 fb $Zb\bar b b\bar b$ process and the $\sim$ 2 pb $Zb\bar
bjj$ process (where $j$ is any non-$b$ quark or gluon, at least two
jets have $p_T>200$ GeV, and the $Z$ decays to neutrinos.)
\footnote{I am grateful to J.R.~Walsh for providing these background
samples, which were simulated using MadEvent \cite{Madevent}.} The
numbers of events shown in the plots are 4200 for $Zb\bar b\bar b$ and
10000 for $Z b\bar b jj$.

For these two backgrounds, and presumably other multi-jet processes
that are the dominant backgrounds remaining after simple cuts, it
appears the transverse thrust variable is not helpful.
Neither signal nor backgrounds resemble back-to-back di-jets, while
neither is spherical, so no cut removes a large fraction of the
background without removing most of the signal.  But the situation
with the cluster mass variable is more promising, as shown in
\reffig{xMclus} for signal and in 
\reffig{xMcluszj4} for the two backgrounds.  The cluster masses are
here constructed from jets, with $p_T>25$ GeV and $|\eta|<2$ to reduce
sensitivity to the underlying event and initial state radiation.
(This is in contrast to \cite{HVWis}, which computed this quantity at
parton-level; because of jet merging in the present signal, such an
approach would not be
reliable here.)  The backgrounds (for which jet-parton correspondence
is more likely to hold) are shown at parton-level; although this
reduces the cluster mass in some events, the effect appears small
enough to not affect the general conclusions below.

The larger $Zb \bar b jj$ background is clearly the more serious
problem.  It can be reduced if three $b$ tags are required of the
events, but at a cost of considerable signal in the A cases.  The
cluster mass distribution of the signal for the A cases lies
underneath the background, and appears not to be useful.  By contrast, the B
cases (which, as in \cite{HVWis}, have few invisible final-state
particles) are much more forgiving, as they are for many variables;
the cluster invariant mass moves the signal far from the backgrounds
shown.  A loose cut on this variable, combined with other selection
criteria (such as at least three $b$-tagged jets) should help this
signal to stand out.

As an aside, note that the v-pion mass appears visibly in the cluster
mass distribution.  There is a small but non-negligible probability
that a one or both hemicylinders contains only a single v-pion, so
that the cluster mass is just the v-pion mass.  However the region in
which this is easiest to see lies underneath the background.
It seems unlikely that this fact can assist with 
identifying the signal.

As noted in \cite{HVWis}, the cluster mass variable is not
sufficiently robust for use in a realistic analysis.  Studies for the
present paper have shown marked dependence, at the level of 20 percent
or more, on the treatment of the underlying event and initial state
radiation.  The plots above should therefore be treated with caution.
They are useful for characterizing differences between signal and
background, but a more stable version of this variable should be used
in any experimental analysis.

The conclusions in this section are thus preliminary.  More robust
event-shape variables should be studied, but the cluster mass appears
to reduce background in the B cases.  However, the B cases are already
distinctive in other ways.  This variable may be most useful in case
B3, where the number of jets is not so extreme as in case B1, but the
total invariant mass of the jets in each hemisphere is still very
large.  On the other hand, this particular variable may not be so useful
in the A cases, so a
different event-shape variable must be sought.

\section{Clinching the Case: Detecting the v-pion resonance}
\label{sec:resonance}

I have discussed a number of features of the signal which make its
phenomenology atypical.  Despite the unusual features of the signal
outlined above, they are not obviously sufficient to allow for easy
separation of signal from background, if the signal
cross-section is indeed 10-100 fb.  The standard model
backgrounds are large and variegated, consisting of tails of
distributions from a number of different processes.  A complete and
convincing study will be difficult with present tools.  In any case,
it seems unlikely that the backgrounds can be understood well enough
from data to allow a counting experiment, especially in the A cases,
where the number of jets, tracks, vertices, {\it etc.} is not so large.
\makeAfig{.336}{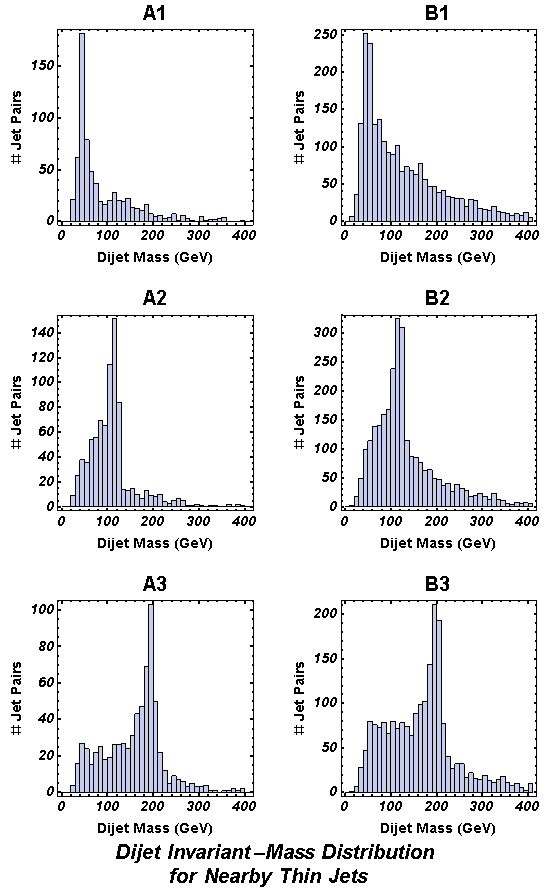}{For the case studies, over 1000
events, the distribution of dijet invariant mass for pairs of 
thin jets ($m_j<0.15 p_T$, $p_T>25$ GeV, $|\eta|<3$)
that are nearby ($\Delta R <1.2$).}  
\makeAfig{.336}{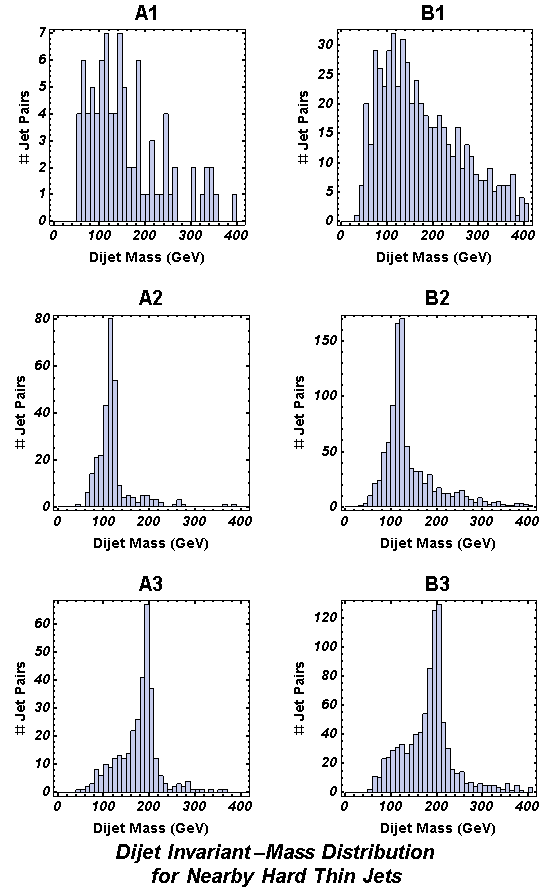}{As in
the previous figure, except that the jets have $p_T>100$ GeV
and pairs are within $\Delta R = 0.9$ of one another.}

Instead, it seems likely that a different strategy is needed.  Given
the unusual features of the signal discussed above, one might apply
loose cuts on these features that have high efficiency for the signal.
(For instance, one could require substantial \MET\ and/or $\hat H_T$, 
several jets with at least three tagged, indications of many displaced tracks
and vertices, {\it etc.})
The standard model background surviving the
cuts would not be calculable or easily measured, so the
signal-to-background ratio would not be well-known.  But within this
enriched sample, one could then search for the key kinematic feature
of the signal -- the v-pion resonance -- which if observed would
confirm that new physics is present.


Simply plotting dijet invariant masses, where the jets are
selected at random, cannot reveal the v-pion resonance.
The huge combinatoric background, the fact that many jets contain
multiple $b$-quarks, and relatively poor resolution for jet momentum
and energy would eliminate any signal.  

Since many v-pions are boosted, they often form a single jet, or
dissociate into two nearby jets.  I will use these facts below in
studying both dijet invariant mass $m_{jj}$ and single-jet invariant
mass $m_j$ below.  Clearly both might be used; there is no sharp
dividing line between them in any case, since any jet algorithm
separates the two in an arbitrary way.  (Approaches that avoid this
division are under study \cite{mptstudy}.)  It appears that it is
important to choose one's jets carefully.

A proper study of these variables would account for the finite
resolution in jet energy, momentum and mass.  There are many issues
here, some beyond the scope of a theoretical investigation.  In this
study I will include decays, showering and hadronization
but will work only with a perfect calorimeter -- perfect except
for its granularity of 0.1$\times$ 0.1 in $\eta$ and $\phi$ and its
limited pseudorapidity coverage --- and will show that substantial
challenges arise even before realistic detector issues are accounted
for. 
\makeAfig{.366}{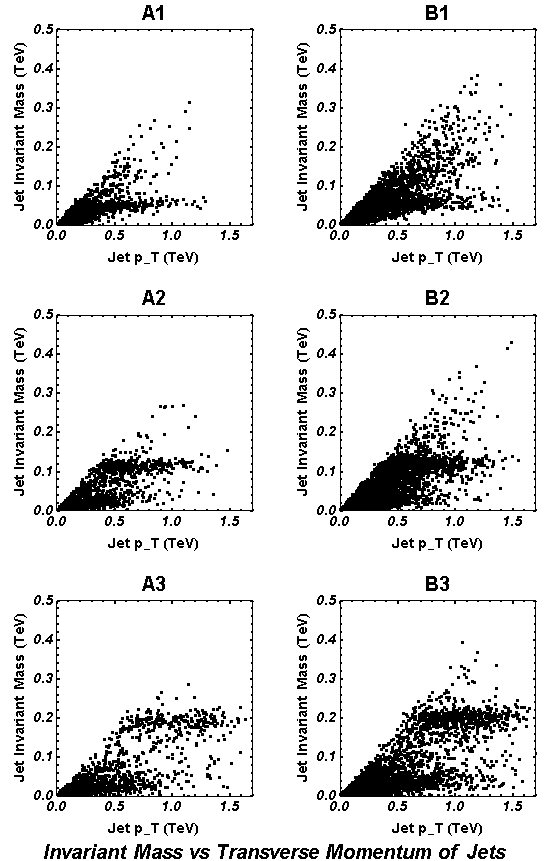}{For jets with $p_T>20$ GeV, the
distribution of $m_j$ versus $p_T$, in TeV.}
\makeAfig{.316}{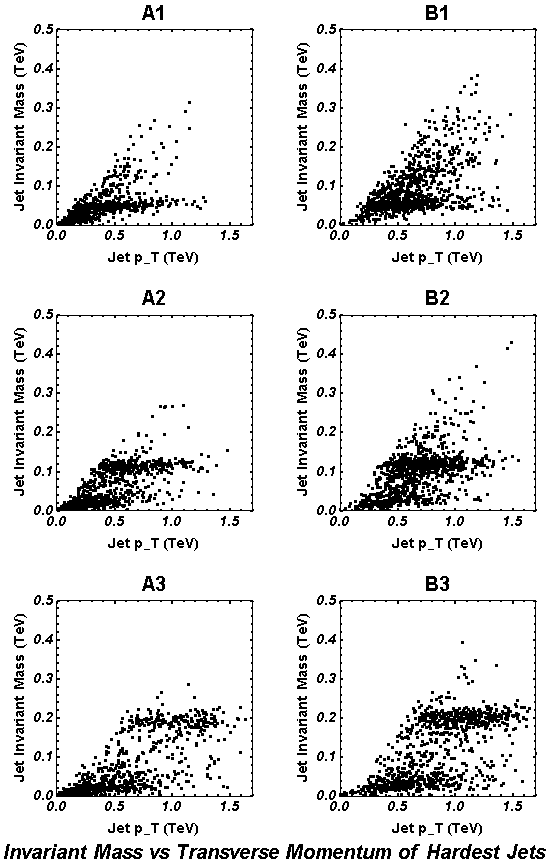}{For the highest-$p_T$ jet in each event,
the distribution of $m_j$ versus $p_T$, in TeV.}

\subsection{Dijet masses}
\label{subsec:dijet}

Let us begin with dijet invariant mass.  A plot
of the invariant masses of all pairs of jets above a certain $p_T$ cut would
suffer from an overwhelming combinatoric background, because of the
high jet multiplicity.  Instead, it is best to use the fact that
high-energy $Z'$ decays provide a substantial boost to many of the
v-hadrons, as we saw in Sec.~\ref{sec:casestudies}.  Most high-$p_T$
jets are either single v-hadrons, whose decay products have merged, or
they represent a single quark produced by the decay of a v-hadron
whose other decay product will lie close by in $\Delta R$.  The
difference between these two cases is (on average) that the former
class of jets will have a single-jet invariant mass $m_j$ larger than the
latter class.  In particular, we will distinguish between ``thin'' and ``thick'' jets, thin jets being those with $m_j<0.15 p_T$, and thick jets
being those with $m_j>0.15 p_T$.  
(This terminology has been introduced in
\cite{mptstudy}.)  
By selecting thin jets with
high $p_T$, and plotting the dijet invariant mass of pairs of such jets
with $\Delta R$ not large, one might hope to significantly reduce the
combinatoric background.  (This same method works for finding
$W$ bosons in high-energy $t\bar t$ events.)

In \reffig{xjjmassA} is shown a plot of dijet masses for cone jets
(defined as described in Sec.~\ref{subsec:jets} and in Appendix
\ref{app:jets}.)  I demand that both jets are thin and have $p_T>25$ GeV and
$|\eta|<3$, and require that the two jets have $\Delta R<1.2$.
In \reffig{xjjmassB} is shown a similar plot with lower statistics but
lower background, using $p_T>100$ GeV and $\Delta R<0.9$.  Either
approach is a challenge for cases A1 and B1, which is not surprising,
since reconstruction of a 50 GeV resonance using jets is no easy task.
As can be seen from the plots, the number of v-pions reconstructed is
disappointingly low; recall there are thousands in the data (see Table
\ref{table:cases}.)  Smearing and mismeasurements, not included here,
will only make matters worse.  It would appear that dijet invariant
mass is not a particularly good variable for reconstructing the v-pion
resonance.
\makeFfig{.47}{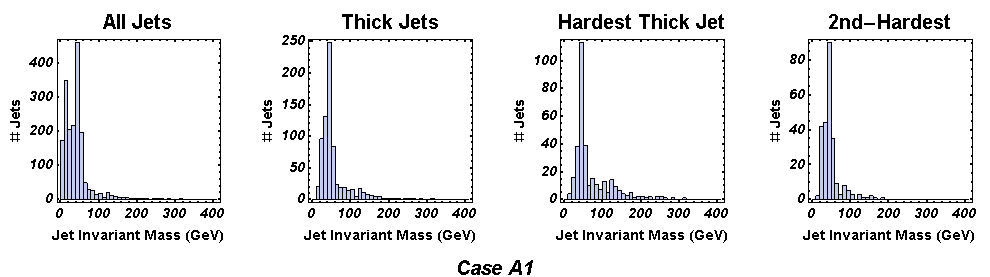}{For the case studies, over 1000
events, the plots, from left to right, show (a) the distribution of
single jet invariant mass $m_j$ for jets with $p_T>100$ GeV and
$|\eta|<3$; (b) the same, with the additional condition that the jets
be ``thick'', namely $m_j/p_T>0.15$; (c) the invariant mass of the
highest-$p_T$ jet (if $p_T>100$ GeV, $|\eta|<3$, and $m_j/p_T>0.15$),
and (d) the same as the previous plot but for the second-highest-$p_T$
jet (with the same cuts.)}

\subsection{Single jet masses}
\label{subsec:jetmass}

Now let us turn to single jet invariant mass $m_j$.  In
\reffig{xpTmasshj} $m_j$ versus $p_T$ is shown for all jets in these
events with $|\eta|<3$.  The same is shown in \reffig{xpTmasshij} for
the highest-$p_T$ jet in each event.  In both classes of plots, a band
of jets with invariant mass near to the v-pion mass is clearly seen.
However, this information is not entirely observable.  High $p_T$ QCD
jets will often develop an invariant mass of order 15 percent of their
$p_T$ just from the emission of a moderate-$k_T$ gluon that lies
outside the parton-shower approximation used in PYTHIA's simulation of
jets.  Also, jets which are too narrow will not have a well-measured
invariant mass, for various detector-related reasons.  One might
therefore wisely exclude from analysis any jet whose mass is less
than, say, 15 or 20 percent of its $p_T$, thus excluding the hardest
jets.  
Refinements of this measurement deserve more attention than can be
given here.  However, some preliminary indications are presented in
\reffig{xhijetmas}, where the single jet invariant mass of all jets
with $p_T>100$ GeV is shown on the left, and the same with the
additional condition that the jet be ``thick'', $m_j>0.15 p_T$, is
shown in the second-to-left column.  Clearly the additional condition
has the advantage of removing many ordinary isolated jets, reducing
the QCD continuum that peaks at low mass, and allowing the signal to
stand out more clearly.  This more refined measure of invariant mass
is shown again in the right-hand plots of \reffig{xhijetmas}, for the
highest-$p_T$ and second-highest-$p_T$ jet in each event.
Interestingly, in case B1 --- the case with highest multiplicity,
where the highest-$p_T$ jet often is a merging of more than two
partons (\reffig{xpINj}) --- the second-highest-$p_T$ jet shows the
v-pion mass more clearly.  The correlation between the masses of the
two highest jets can also be a useful variable; a scatter plot of the
masses of the two highest-$p_T$ jets (if both are central and thick)
is shown in \reffig{xmassj2vj1}.  Note that in case B1 the high
multiplicity tends to spread out the peak, whereas in case A3 the peak
is almost invisible due to low statistics, but for the other cases
this plot can help reveal the resonance.
\makeAfig{.366}{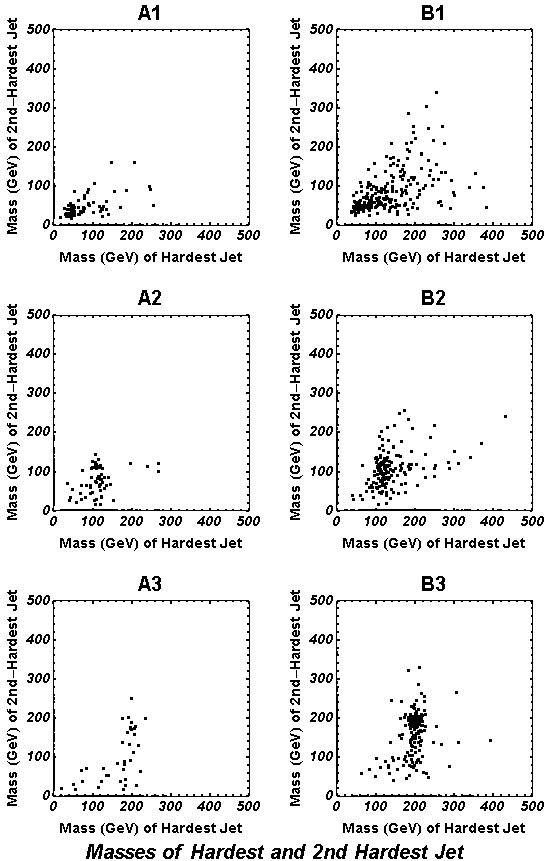}{Jet-invariant mass (in GeV) for the
highest and second-highest $p_T$ jets in each event; jets
are required to have $p_T>100$ GeV, $m_j>0.15 p_T$, and $|\eta|<3$.}

The fact that the single-jet invariant mass is a good observable in
all 6 cases is a simple consequence of the fact that relatively light
particles are being produced with a large boost, through the decay of
a heavy $Z'$.  (Clearly the same strategy will not work for v-pions produced in
decays of lighter particles, such as Higgs bosons \cite{HV1,HV2} or
supersymmetric particles \cite{HV3}.)
There are important and little-studied
backgrounds from all-hadronic decays of boosted $W$'s, $Z$'s, and $t$'s,
which are produced with an enormous rate.  These will swamp 
any new resonance unless events are first selected
with the unusual features of this signal. 
Should the v-pion mass lie close to 80-90 GeV or to 170
GeV, the difficulties will be very much greater.  We must hope
nature does not choose this scenario, or at least provides a large
cross-section in return.
\makeAfig{.316}{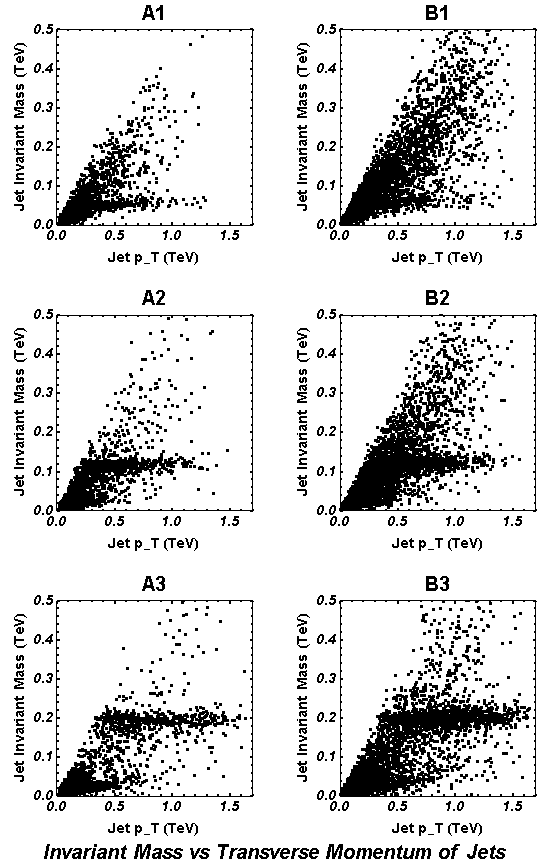}{As in \reffig{xpTmasshj}, but with
cone size 0.7 instead 0.4; 
for jets with $p_T>20$ GeV, the
distribution of $m_j$ versus $p_T$, in TeV.}

\subsection{Some Improvement Using Fatter Jets}
\label{subsec:fatjets}

We have seen that single jet invariant mass reconstructs more v-pions
than does dijet invariant mass.  By widening the jet cone, one might
hope to reconstruct even more.  Here I will explore 
increasing the cone radius from 0.4 to 0.7, and will show
some increase in efficiency. 
\makeFfig{.47}{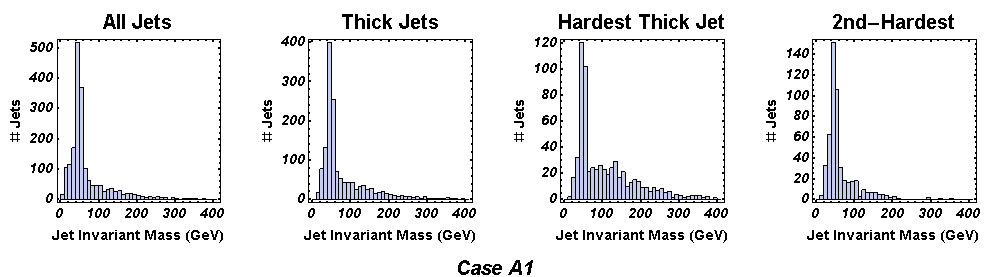}{As in \reffig{xhijetmas}, but with
cone size 0.7 instead 0.4; from left to right, (a) the mass $m_j$ of
jets with $p_T>100$ GeV and $|\eta|<3$; (b) the same but with $m_j/p_T>0.15$;
(c) $m_j$ for the highest-$p_T$ jet with
jets with $p_T>100$ GeV, $|\eta|<3$ and $m_j/p_T>0.15$; and (d), the same but for the second-highest-$p_T$ jet.}
\makeAfig{.366}{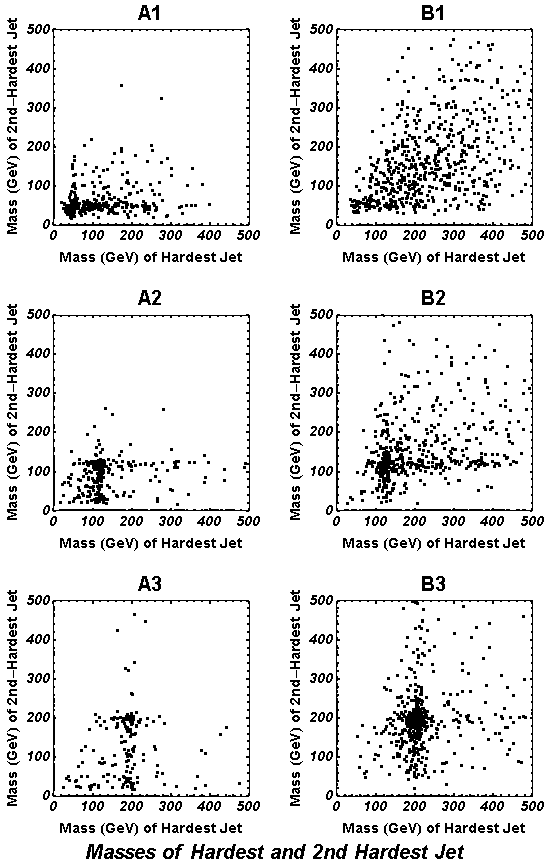}{As in \reffig{xmassj2vj1}, but with
cone size 0.7 instead 0.4; jet-invariant mass (in GeV) for the highest
and second-highest $p_T$ jets in each event, where jets are
required to have $p_T>100$ GeV, $m_j>0.15 p_T$, and $|\eta|<3$.}

Comparing \reffig{xpTmasshj7} to \reffig{xpTmasshj}, we see that a
larger fraction of the jets of radius 0.7 are single v-pions, compared
to those of radius 0.4.  This is very clear in the left-hand plots of
\reffig{xhijetmas7}; whereas the thickness criterion (the cut in
$m_j/p_T$) is essential to remove random jets in the signal in
\reffig{xhijetmas}, it is less essential, but still effective, with the
larger cone size.  Comparison of \reffig{xhijetmas7} with
\reffig{xhijetmas}, and of \reffig{xmassj2vj17} with
\reffig{xmassj2vj1}, shows that the larger cone size generally allows
a marked improvement in both efficiency and resolution, for both the
hardest and second-hardest jet in the event.

The one interesting exception, among the case studies, is case B1.
Here the number of partons in the final state is so large that
confusion background dominates.  Most high-$p_T$ jets contain multiple
$b$ quarks, and the hardest jets tend to contain more than 2, a
tendency already visible in \reffig{xpINj}.  With a cone size of 0.4,
rather few of the high-$p_T$ jets are single v-pions, and a larger
cone-size makes this problem worse.  It appears it is best in this
case to work with the jets of radius 0.7 that do not have the highest
$p_T$, and even then the background from random jets in the signal is
rather large (see \reffig{xhijetmas7}, fourth line, second plot from left).
Background from standard model processes would be very problematic
except for the fact that case B1 is also the easiest case to separate
from the standard model using other methods.  This case has the
highest multiplicity of jets ($\sim 7$) and vertices ($\sim 20$), the
highest multiplicity of secondary muons ($\sim 3$), the most tracks
($\sim 100$) and displaced tracks ($\sim 50$), very high $\hat H_T$
($\sim 1.6$ TeV), and a striking event shape (sum of the two cluster
masses $\sim 1.2$ TeV).  Perhaps this signal can even be identified in
a counting experiment, where the standard model backgrounds can be
estimated from the data by looking at event samples that share some
but not all of these striking features.  It is conceivable that the v-pion
resonance can be better identified with a more sophisticated variable
than single jet mass, looking more carefully at the substructure of
the jets.  (It is even possible that, with so many v-pions per event,
and with a bit more statistics than available here, the v-pion can be
discovered through its rare tree-level decay to muon pairs or its
loop-induced decay to photon pairs.)  More generally, it is important
to study further how best to look for resonances in
very-high-multiplicity signals, such as case B1.

\subsection{Comments}

I have given evidence that single jet mass, using a larger cone size
than typically used for jets at the LHC, is the variable to use in
searching for the v-pion resonance.  Why are the fatter jets a better
choice?  If a single boosted v-pion has a boost factor greater than
6, it will typically form a single {\it thin} jet.  
Thus, thick jets arise from v-pions with a boost below 6, whose
daughters typically have an opening angle of order 0.3
or larger.  This is why, if a thickness criterion is applied, jets
of radius 0.7 have a much higher efficiency for containing a single
v-pion than jets of radius 0.4.

These conclusions are somewhat suspect, and the plots above
unrealistic, because no energy smearing or magnetic field were
included in the simulation of the calorimeter. A more serious study is
needed, using a more complete detector simulation.  Several remarks
are in order.

It will be of considerable interest to learn the jet-mass resolution
of the LHC detectors.  This information should be available in the
early data from studies of $W$ bosons and $t$ quarks in
boosted $t\bar t$ events.  It will also be interesting to see plots of
the QCD continuum background to jet-mass measurements.

Since angular resolution is essential for jet mass measurements, the
curvature of tracks due to the magnetic field should be 
removed.  Clearly the use of something like ``particle flow'' ---
combining angular information from the tracker as well as
the electromagnetic and hadronic calorimeters --- should significantly
aid in improving the resolution on jet mass.  

An additional handle for reducing backgrounds to boosted v-pions may
lie in detecting the substructure within the jet, such as recently
considered in \cite{boost}.  This is easiest to do with clustering
algorithms such as the $k_T$ algorithm, and using tracking as well as
calorimetry.  In fact, for the most energetic v-pions, whose daughters
have the lowest angular separation, the granularity of the
calorimeter, especially its hadronic component, will badly degrade
jet-mass resolution.  In this case tracking is crucial (see for
example \cite{phenotalk}.)  By studying the locations of the
highest-$p_T$ tracks, one may be able to significantly improve angular
resolution on the substructure of the jet, and improve the invariant
mass measurement.  These issues deserve much more attention and will
be explored elsewhere \cite{mptstudy}.

\section{Conclusion}

In this paper I have investigated the phenomenology of a large class
of hidden valley models, illustrated by six case studies within a
particular model, but intended to apply in a much broader context.
The models present a single new narrow resonance (more precisely, in
the particular case studied, a triplet of nearly degenerate resonances
$\pi_v^0,\pip,\pim$) that decays promptly, and predominantly to heavy flavor.
Importantly, there is no measurable rate for decay to $\mu^+\mu^-$ or
$e^+e^-$.  In this sense this class is largely orthogonal to the large
class studied in \cite{HVWis}.  In production mechanisms ($Z'$ decay)
that lead to high-multiplicity heavy-flavor final states, the
phenomenology is very different from that of most supersymmetry,
technicolor, Randall-Sundrum, or little Higgs models that have
so often been studied.

The six case studies were divided into the A cases (where the $\pipm$
are invisible and stable) and the B cases (where all three types of
$\pi_v$ decay promptly).   The main results were the following:
\begin{itemize}
\item Missing energy and total transverse energy are useful global variables.
\begin{itemize}
\item In the A cases,  \MET\ and $\hat H_T$ are of the same
order, many hundreds of GeV for 50 GeV v-pions and closer to a TeV for
200 GeV v-pions.
\item In the B cases, the \MET\ is much smaller and the $\hat H_T$ is
in the 1.5--2.5 TeV range.
\end{itemize}
\item Multiplicities of v-hadrons,
and their daughter standard model partons (mostly bottom quarks),
are very large.
\begin{itemize}
\item In the A cases, the number of v-hadron daughters ranges from an average of
3 with a tail to 12 (for heavy v-pions) to an average of 8 with a tail
to 24 (for lighter v-pions).
\item In the B cases, the number of v-hadron daughters ranges from an average
of 8 with a tail to 20 (for
heavy v-pions) to an average of 20 with a tail to 42 (for lighter v-pions).
\end{itemize}
\item However the multiplicity of jets is somewhat smaller, because jets often
contain multiple v-hadron daughters; this is due to the high boost and high
concentration of the v-pions.  
\begin{itemize}
\item In the A cases, this effect is especially important, because by
reducing the number of hard jets it leaves the signal with larger standard
model backgrounds.
\item In the B cases, the effect is more pronounced, but the resulting
multiplicity of jets is still large compared to most standard model
processes.
\end{itemize}
\item In the A cases, counting of heavy-flavor-tagged jets
appears insufficient for separating signal from background, because the
multiplicity of jets is too low.  The situation in the B cases is 
somewhat more
promising.
\item The signal is particularly unusual in the numbers of vertices,
of tracks, and of displaced tracks, and the clustering of and
correlations among
jets, tracks and vertices.  These features may be
useful in separating signal from pbackground.
\item Other observables, including event-shape variables (particularly
one similar to the $M_{cluster}$ variable of \cite{HVWis}) and numbers
of secondary muons, might serve as additional tools for event
selection, though their utility needs more study.  They may not be
helpful in the A cases, and they may not be needed in the B cases.
\item If the unusual features of the signal can be used to obtain a
sample with a reasonable signal-to-background ratio, an attempt can be
made to find the v-pion resonance.
\begin{itemize}
\item Because of combinatorics, a naive approach to dijet masses cannot work.
\item Since many v-pions are boosted, it is better to consider dijet
masses of jets which are close in $\eta$ and $\phi$, or single-jet
masses of individual jets which are ``thick'' (have a large
mass-to-$p_T$ ratio.)
\item Single-jet mass for a relatively fat jet definition seems to be
an observable with a high efficiency for the signal.  Here $R=0.7$
cone-jets were shown to be better than those with $R=0.4$, but a
thorough study of the best jet radius was not carried out, nor were
other algorithms carefully considered.  More work
remains to optimize this measurement \cite{mptstudy}.
\item Correlations between the single-jet mass of the highest- and
second-highest-$p_T$ jets may be useful in reducing background.
\item With extreme multiplicity, as in case B1, jet mass for
high-$p_T$ jets is less useful, because too many of these jets contain
more than a single v-pion.  An alternative observable is not yet
known, but also such signals are particularly spectacular, with low
standard model backgrounds, and a discovery claim might not require
reconstruction of the resonance.
\item It is possible but rather unlikely
that multiplicities and event rates will
be high enough to allow reconstruction of the $\pi_v$ resonance
through the rare decays $\pi_v\to \mu^+\mu^-$ or
$\pi_v\to \gamma \gamma$.
\end{itemize}
\end{itemize}
In all cases, the phenomenological issues that arise are somewhat
unusual, and need to be explored further in more fully realistic
studies.  Generating relevant and realistic background
samples for these unusual signals represents a substantial challenge.

I should again emphasize that the issues and observables discussed in this paper
are not limited to these specific models, but will apply more broadly.
In a number of classes of hidden valley models, new resonances that
decay to heavy flavor are produced in high-multiplicity environments,
sometimes with a substantial boost.  This can also happen in models 
beyond the hidden valley scenario.

Conversely, it is important to stress that this particular class of
models is special, and relatively easy, in that there is only one
v-hadron resonance to be found. In many other models, there will be
multiple v-hadrons of different masses, and so the resonance signal
just described will be spread out among several resonances, making it
harder to extract.  One may still hope, however, for an enhancement in
jet and dijet masses that will be inconsistent with Standard Model
background.  And some models are even easier to find, as in the
example \cite{HV1} studied phenomenologically in \cite{HVWis}, because
of the presence of electron-pair and muon-pair resonances.  These may
easily be seen above background even with a rather impure sample of
hidden valley events.

Several other classes of hidden valley models with distinct
phenomenology from those studied here and in \cite{HVWis} remain to be
investigated.  To examine these in detail requires a significant
extension of the current Monte Carlo simulation package.  An initial
extension is now complete \cite{HVMC1}.  Preliminary results on some
models, with their own unusual signals, will be presented soon
\cite{hvstudy2}.

\

I thank G.~Ciapetti, A.~De Roeck, C.~Dionisi, S.D.~Ellis, S.~Giagu,
T.~Han, J.W.~Huston, P.~Loch, H.J.~Lubatti, G.P.~Salam, Z.~Si,
C.K.~Vermilion, J.R.~Walsh and K.M.~Zurek for useful discussions.  I
am especially indebted to S.~Mrenna and P.Z.~Skands for advice and
assistance in the writing of the HVMC 0.5 Monte Carlo package.  This
work was supported in part by Department of Energy grant
DE-FG02-96ER40956.

\appendix

\makeAfig{.301}{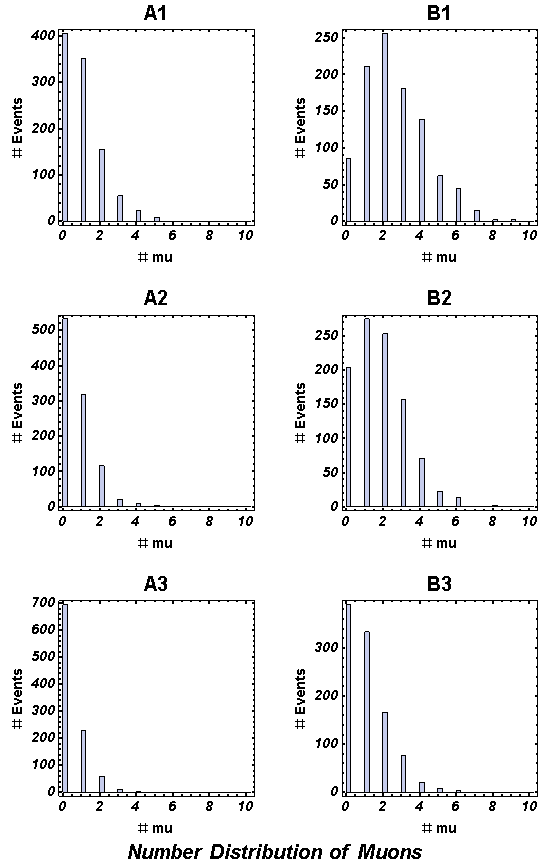}{The distribution of the number of secondary muons.}
\makeAfig{.301}{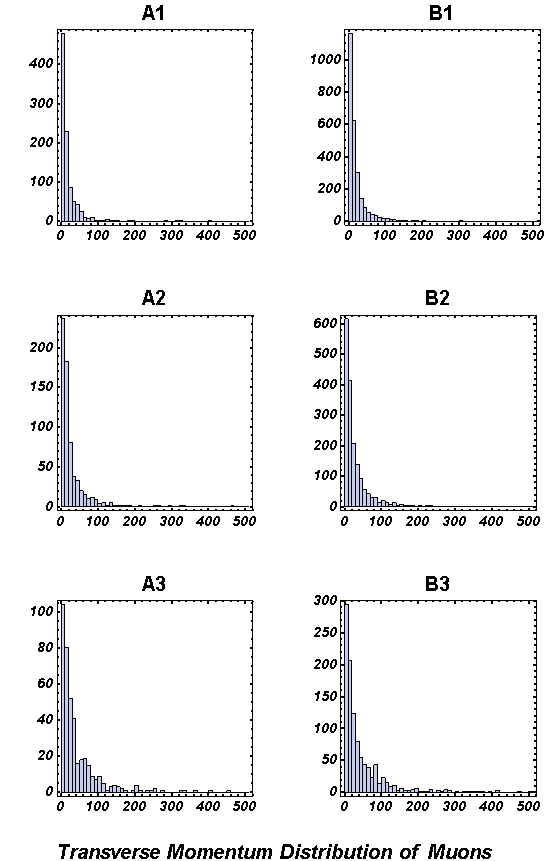}{The $p_T$ distribution of the secondary muons.}

\section{Observations concerning leptons}
\label{app:leptons}

Plots of the
number of central prompt or semi-prompt (appearing within the beampipe)
muons, with $p_T>3$ GeV and
$|\eta|<2$, are shown in \reffig{xNmu}, and their $p_T$ distributions are
shown in \reffig{xpTmu}.  No isolation criteria are imposed. 
\makeAfig{.25}{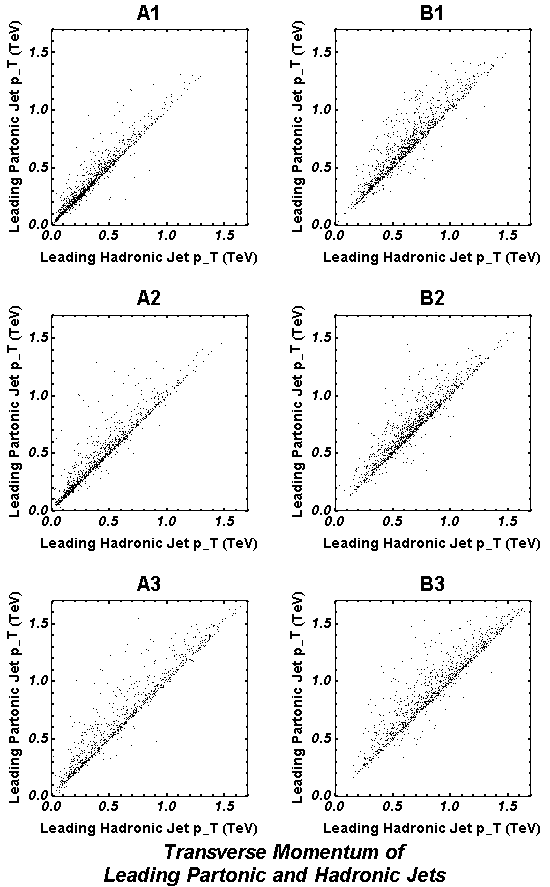}{Correlation between the $p_T$ of the
highest-$p_T$ partonic jet and the $p_T$ of the highest-$p_T$ hadronic
jet, constructed using the midpoint cone algorithm with $R=0.4$ (see
Table \ref{table:algos}.)}
\makeAfig{.25}{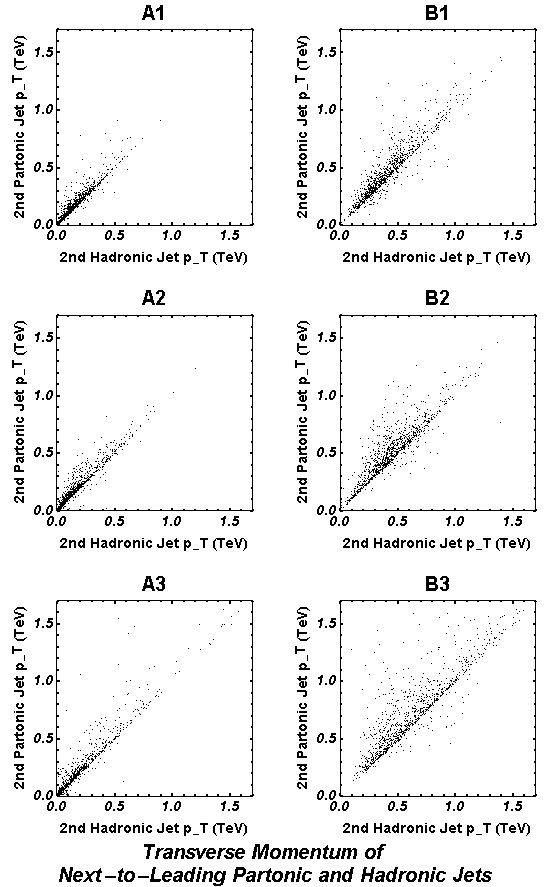}{As in the previous plot, for the
second-highest-$p_T$ partonic and hadronic jets.}
\makeAfig{.25}{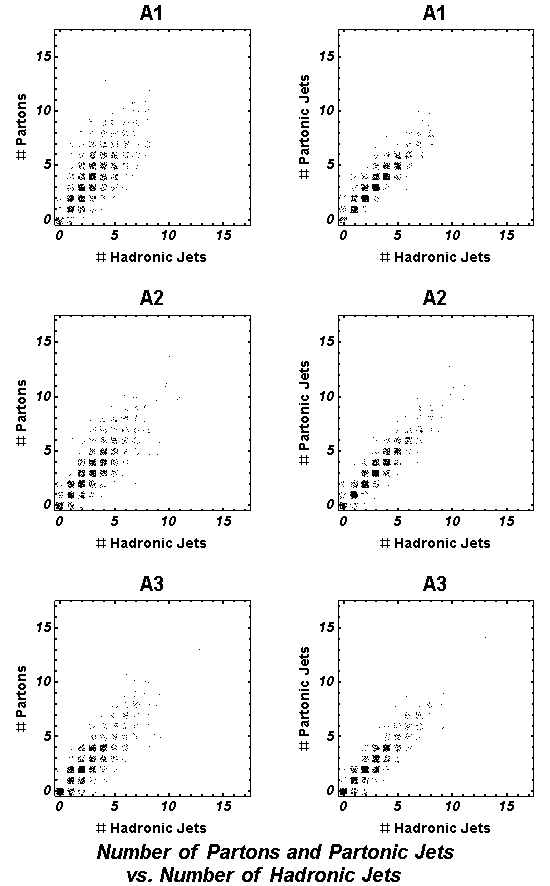}{As in \reffig{xNpNpjA}, for the $k_T$ algorithm with $R$ parameter $0.52$ (see Table \ref{table:algos}.)}

\section{Checking the Correspondence of Hadronic Jets and Partonic Jets}
\label{app:jets}

Here I will show more carefully that the partonic and hadronic
jets, constructed with a reasonable algorithm, do correspond, as claimed
in the main text.

The jet algorithms used in this study are shown in Table
\ref{table:algos}.  All studies use the multi-algorithm software SpartyJet of
\cite{huston}, which includes the FastJet $k_T$ algorithm
\cite{fastjet}.  In the plots within the main text, the midpoint
merging version of the cone algorithm was used, with cone size $R =
0.4$.  In this appendix, a cone of $R=0.7$ is also used, along with the
$k_T$ algorithm with the $R$ parameter set to give $k_T$ jets of
radius approximately 0.4 and 0.7.

First, let us check the correspondance between
hadronic jets and parton jets.  One may confirm that the $p_T$ of the
highest-$p_T$ partonic jet matches with the $p_T$ of the highest-$p_T$
hadronic jet; this is shown in \reffig{xpT1pjhj}.  The psedorapidities
match as well.  Nor is this an accident limited to the highest-$p_T$
jet; the same figure for the second-highest $p_T$ jet shows that the
strong correlation between the hadronic and partonic jets persists.

\begin{table}[b]
\begin{tabular}[c]{|c|c|c|c|}\hline
\multicolumn{4}{|c|}{Midpoint Cone Algorithm} \\ \hline 
 cone & seed       &  search cone & merge   \\
radius & threshold & area fraction & fraction  \\
 \hline 
0.4 & 1 GeV & 0.25 & 0.75 \\
\hline
 0.7 & 1 GeV & 0.25 & 0.75 \\ \hline \hline
\multicolumn{4}{|c|}{FastJet $k_T$ Algorithm} \\ \hline 
 R         & minimum    &   & \\
 parameter & cell $p_T$ &  $d_{cut}$        &          \\
 \hline 
0.52 & 5 GeV & (5 GeV)$^2$ & \\
\hline
0.91 & 5 GeV & (5 GeV)$^2$ & \\
\hline
\end{tabular}
\caption{The parameters used in the cone and $k_T$ algorithms used
in this study.}
\label{table:algos}
\end{table}
\makeAfig{.25}{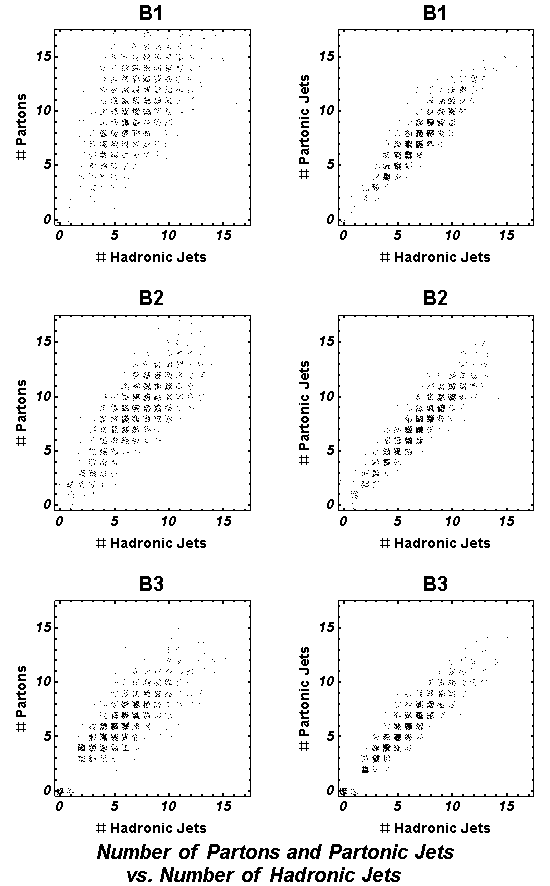}{As in \reffig{xNpNpjB}, for the $k_T$ algorithm with $R$ parameter $0.52$ (see Table \ref{table:algos}.)}

Now I turn to the $k_T$ algorithm.  As with the cone algorithm we see,
in Figs.~\ref{fig:kNpNpjA} and \ref{fig:kNpNpjB}, a failure of the
correspondence of partons to hadronic jets, and better agreement
between partonic jets and hadronic jets.  Figures \ref{fig:kpT1pjhj}
and \ref{fig:kpT2pjhj} further confirm the quantitative correspondence
of the partonic and hadronic jets.  Thus the cone and $k_T$ algorithms
both allow reconstruction of the partonic jets using hadronic jets.
\makeAfig{.25}{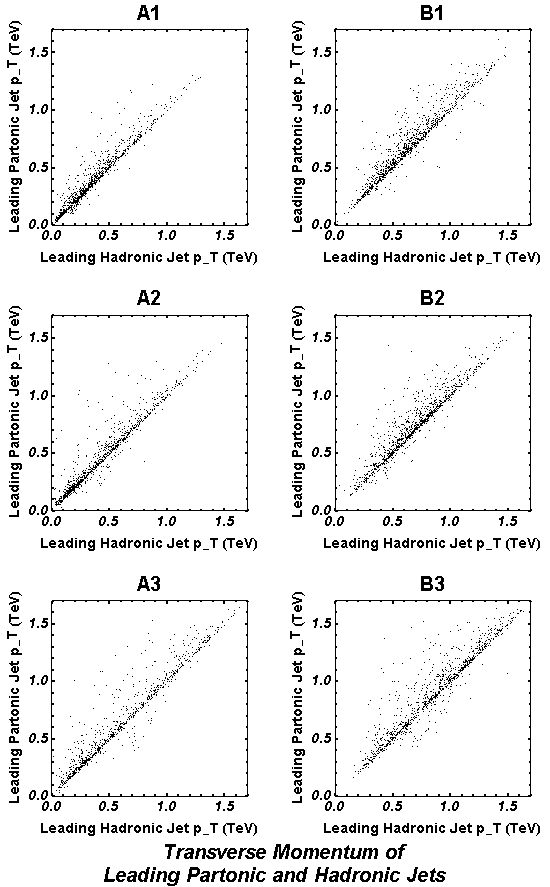}{As in \reffig{xpT1pjhj}, for the $k_T$
algorithm with $R$ parameter $0.52$ (see Table \ref{table:algos}.)}
\makeAfig{.25}{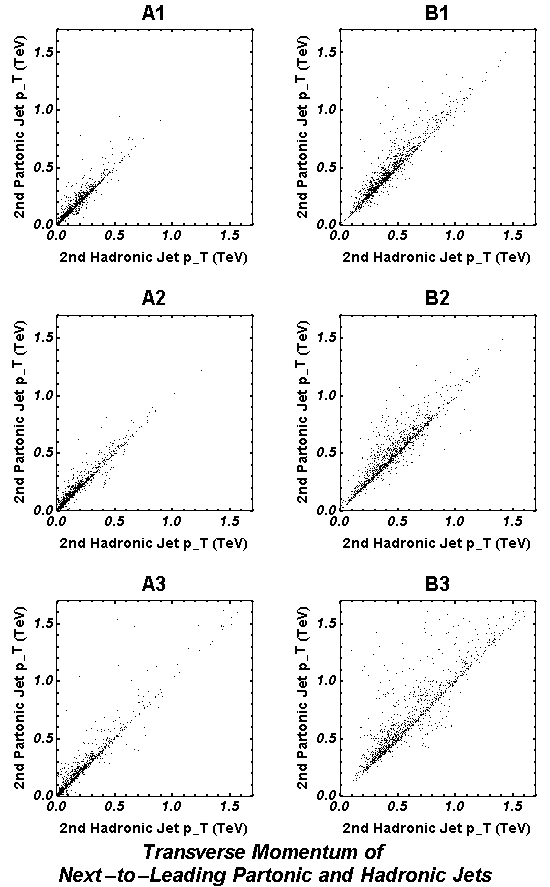}{As in \reffig{xpT2pjhj}, for the $k_T$
algorithm with $R$ parameter $0.52$ (see Table \ref{table:algos}.)}

These studies have been repeated for jets with larger settings of the
cone size or $R$ parameter (0.7 and 1.0 for cone jets).  The
correspondence of partonic and hadronic jets continues to hold firm.

\end{document}